\documentstyle[prl,aps,epsfig]{revtex} % 1 col mode

\tighten
\begin{document}
%
%\baselineskip=8.5mm  % preprint mode
%\draft               % preprint mode
% 2 col mode:
%\twocolumn[\hsize\textwidth\columnwidth\hsize\csname@twocolumnfalse\endcsname

% define commands for international characters
\catcode`\ä = \active \catcode`\ö = \active \catcode`\ü = \active \catcode`\Ä = \active
\catcode`\Ö = \active \catcode`\Ü = \active \catcode`\ß = \active \catcode`\é = \active
\catcode`\è = \active \catcode`\ë = \active \catcode`\ô = \active \catcode`\ê = \active
\catcode`\ø = \active \catcode`\ò = \active
\defä{\"a} \defö{\"o} \defü{\"u} \defÄ{\"A} \defÖ{\"O} \defÜ{\"U} \defß{\ss} \defé{\'{e}}
\defè{\`{e}} \defë{\"{e}} \defô{\^{o}} \defê{\^{e}} \defø{\o} \defò{\`{o}}

\title{Collective enhancement and suppression in Bose-Einstein condensates}

\author{Wolfgang Ketterle and Shin Inouye}
\address{Department of Physics and Research Laboratory of Electronics, \\
Massachusetts Institute of Technology, Cambridge, MA 02139, USA}

\date{\today}
\maketitle

\begin{abstract}
The coherent and collective nature of a Bose-Einstein condensate
can enhance or suppress physical processes.  Bosonic stimulation
enhances scattering in already occupied states which leads to
matter wave amplification, and the suppression of dissipation
leads to superfluidity. In this summer school notes we present
several experiments where enhancement and suppression have been
observed and discuss the common roots of and differences between
these phenomena.
\end{abstract}

\vskip1pc
\tableofcontents
\newpage

When a gas of bosonic atoms is cooled below the transition temperature of Bose-Einstein
condensation, it profoundly changes its properties.  The appearance of a macroscopically occupied
quantum state leads to a variety of new phenomena which set quantum fluids apart from all other
substances.  Fritz London even called them the fourth state of matter~\cite{lond64}.

Many of the key concepts in quantum fluids were derived from studying the weakly interacting Bose
gas, for which rigorous theoretical treatments were possible \cite{huan87,bogo47}. In 1995, with
the discovery of BEC in a dilute gas of  alkali atoms \cite{ande95,davi95evap,brad95bec}, it
became possible to study such a system experimentally. Since then, the interplay between theory
and experiment has been remarkable \cite{dalf99rmp}.

This paper is the fourth major review paper of our group which describes the new techniques and the
physics of Bose-Einstein condensation \cite{kett96evap,kett99var,stamp00leshouches}. These review
papers together give a comprehensive overview of the topics to which our group has contributed.
Each contribution is self-contained, but we avoided major overlap with previous review papers. The
topic of these Cargese summer school notes is enhancement and suppression of physical processes in
a Bose-Einstein condensate.

Many phenomena in Bose condensates involve such enhancement or
suppression. Our recent experiments include the enhancement and
suppression of elastic collisions of impurity atoms
\cite{chik00}, the suppression of dissipation due to
superfluidity \cite{rama99,onof00sup}, and the enhancement
\cite{inou99super} and suppression \cite{stam99phon} of light
scattering.  The common discussion of these phenomena leads to a
better understanding of the underlying principles.  We draw
analogies between light scattering and particle scattering,
between microscopic and macroscopic superfluidity.  We show that
a condensate responds very differently to two different ways of
momentum transfer, light scattering and spontaneous emission
\cite{gorl00spont}. We discuss light scattering in both the linear
and nonlinear regime where bosonically enhanced Rayleigh
scattering led to the amplification of either atoms
\cite{inou99mwa} or light \cite{inou00slow} and work out the
relationship between these two processes.  Finally, the section
on matter wave amplification of fermions discusses the relevance
of symmetry and long coherence time and its relation to quantum
statistics.

\section{Scattering of light and massive particles}
\label{sec:scattering}

Before we discuss light scattering and collisions in a BEC, we want to derive some simple general
expressions based on Fermi's golden rule.  This will help to see the similarities and differences
between the various processes.  When a condensate scatters a photon or material particle, the
scattering is described by the Hamiltonian
\begin{equation}
    \label{eq:hamil}
    {\cal H^\prime} = C \sum_{k,l,m,n} \hat{c}^\dagger_{l}
    \hat{a}^\dagger_n \hat{c}_k \hat{a}_m \delta_{l+n-k-m}.
\end{equation}
Here $\hat{c}_k$ ($\hat{c}^\dagger_k$) is the destruction
(creation) operator for the scattered particles (which can be
photons or massive particles), and $\hat{a}_k$
($\hat{a}^\dagger_k$) is the destruction (creation) operator for
atomic plane waves of wavevector $\bf k$ (free particle states).
The strength of the coupling is parametrized by the coefficient
$C$ (which in general may depend on the momentum transfer), and
the $\delta$ function guarantees momentum conservation (with an
uncertainty of $\hbar/D$, where $D$ is the dimension of the
condensate).

We consider the scattering process where a system with $N_0$
atoms in the condensate ground state and $N_q$ quasi-particles
with wavevector $\bf q$ scatters particles with incident
wavevector $\bf k$ into a state with wavevector $\bf k - q$.  The
initial and final states are\footnote{This choice of final states
implies that we neglect scattering between quasi-particles and
consider only processes involving the macroscopically occupied
zero-momentum state of the condensate. Formally, we replace the
Hamiltonian (Eq.\ \ref{eq:hamil}) by $C \sum_{k,q}
(\hat{c}^\dagger_{k-q}\hat{a}^\dagger_q \hat{c}_k \hat{a}_0 +
\hat{c}^\dagger_{k-q} \hat{a}^\dagger_0 \hat{c}_k \hat{a}_{-q})$.}
\begin{eqnarray}
    \label{eq:states}
    |i\rangle &=&  |n_k, n_{k-q} ; N_0, N_q\rangle  \nonumber \\
    |f\rangle &=&  |n_k-1, n_{k-q} +1 ; N_0-1, N_q+1\rangle
\end{eqnarray}
respectively, where $n_k$ denotes the population of scattering particles with wavevector $\bf k$.
It should be emphasized that, due to the interatomic interactions, the quasi-particles with
occupation $N_q$ are not the plane waves created by the operator $\hat{a}^\dagger_q$, but the
quanta of collective excitations with wavevector $\bf q$.

The square of the matrix element $M_1$ between the initial and final state is
\begin{eqnarray}
    \label{eq:matrix-el1}
    &&|M_1|^2 = |\langle f|{\cal H^\prime}|i \rangle|^2 \nonumber\\
    &&= |C|^2 |\langle N_0=N-1,N_q=1|\hat{\rho}^\dagger({\bf q})|N_0=N,N_q=0
    \rangle|^2 (N_q+1) (n_{k-q} +1) n_k
\end{eqnarray}
where $\hat{\rho}^{\dagger}({\bf q}) = \sum_m
\hat{a}^\dagger_{m+q} \hat{a}_m$ is the Fourier transform of the
atomic density operator at wavevector {\bf q}. The  static
structure factor of the condensate is
\begin{equation}
    S(q) = \langle g | \hat{\rho}({\bf q})
    \hat{\rho}^\dagger({\bf q}) | g \rangle/N
    \label{eq:def_s_q}
\end{equation}
where $| g \rangle=|N_0=N, N_q=0 \rangle$ is the BEC ground state
and $N$ is the total number of atoms. We then obtain for the
scattering matrix element $M_1$
\begin{equation}\label{eq:matrix-el2}
    |M_1|^2 = |C|^2 S(q) (N_q+1) (n_{k-q} +1) N_0 n_k.
\end{equation}

The scattering rate $W_1$ for the process $|n_k, n_{k-q} ; N_0, N_q\rangle \rightarrow | n_k-1,
n_{k-q} +1 ; N_0-1, N_q+1 \rangle$ follows from Fermi's golden rule as
\begin{equation}
    \label{eq:rate}
    W_1 = \frac{2 \pi}{\hbar} |M_1|^2  \delta(E_k - E_{k-q} - \hbar \omega_q^B)
\end{equation}
where $E_k$ is the energy of the incident particle with wavevector
$\bf k$, and $\hbar \omega_q^B$ is the energy of quasi-particles
with wavevector $\bf q$ (which we will later obtain from
Bogoliubov theory). To obtain the net growth rate of $N_q$, one
has to include the reverse process  $|n_k, n_{k-q} ; N_0,
N_q\rangle \rightarrow |n_k+1, n_{k-q}-1; N_0+1, N_q-1 \rangle$
by which atoms scatter \emph{back} into the condensate. The
square of the matrix element $M_2$ for this process is $|C|^2
S(q) N_q n_{k-q} (N_0+1) (n_k+1)$.   The \emph{net} rate $W_+$ of
scattering atoms from the condensate into the quasi-particle mode
$\bf q$ is the difference of the two partial rates $W_+=W_1-W_2$.
Assuming $N_0\gg 1$ (i.e.,\ $N_0+1 \approx N_0)$, we obtain for
the net rate
\begin{equation}
    \label{eq:net-rate}
    W_+ =\frac{2 \pi}{\hbar} |C|^2 S(q) N_0 [n_k (N_q + n_{k-q} +1)-N_q n_{k-q}]
    \delta(E_k- E_{k-q} - \hbar \omega_q^B).
\end{equation}

For large $n_k$ (e.g.\ a laser beam illuminating the condensate) the term $N_q n_{k-q}$ can be
neglected, and the dominant bosonic stimulation term $(N_q + n_{k-q} +1)$ is approximately
$(\tilde{N}+1)$ with $\tilde{N}=\max(N_q, n_{k-q})$. This illustrates that there is no bosonic
stimulation of the net rate by the \emph{least} populated final state.  With the dynamic structure
factor $S({\bf q},\omega)=S(q) \delta(\omega -\omega_q^B)$ Eq.\ \ref{eq:net-rate} simplifies to
\begin{equation}
    \label{eq:net-rate2}
    W_+=\frac{2 \pi}{\hbar^2} |C|^2 n_k \: S\left({\bf q},(E_k-E_{k-q})/\hbar\right) N_0 (N_q + n_{k-q} +1).
\end{equation}

The rate $W_+$ in Eq.\ \ref{eq:net-rate} is the rate for the
Stokes process where $E_k > E_{k-q}$. Momentum transfer $\hbar
\bf{q}$ to the condensate is also possible as an anti-Stokes
process where a quasi-particle with momentum $- \hbar \bf{q}$ is
scattered into the condensate, and the scattered particle
\emph{gains} energy. The net rate $W_-$ for this process is
obtained in an analogous way as
\begin{equation}
    \label{eq:anti-stokes-rate}
    W_- = \frac{2 \pi}{\hbar} |C|^2 S(q) N_0 [n_k (N_{-q} - n_{k-q})-(N_{-q}+1) n_{k-q}]
    \delta(E_k - E_{k-q} + \hbar \omega_q^B).
\end{equation}

The net scattering rates of Eqs.\ \ref{eq:net-rate} and \ref{eq:anti-stokes-rate} involve the
product of three terms.
\begin{itemize}
    \item The static structure factor $S(q)$ represents the squared matrix element for the condensate to
    absorb momentum $\hbar \bf q$.
    \item The $\delta$ function denotes the density of final states.
    \item The bosonic stimulation term represents stimulation by the occupation in the final state
    either of the scattering particles or the condensate.
\end{itemize}

The interplay of these three terms is responsible for the enhancement and suppression of physical
processes in a condensate.  The properties of the condensate as an intriguing many-body system are
reflected in the structure factor.  In Sect.\ \ref{sec:Bragg_spect}, we discuss its measurement
through stimulated light scattering.  The density of states is responsible for superfluidity
because it vanishes for initial velocities of the incident particles which are smaller than the
Landau critical velocity (Sects.\ \ref{sec:collisions} and \ref{sec:macroscopic_flow}). Finally,
bosonic stimulation by the occupancy $N_q$ of final states was responsible for superradiance,
matter wave amplification and optical amplification in a condensate (Sects.\ \ref{sec:SR_and_MWA}
and \ref{sec:Amplification of light in a dressed condensate}).

Sects.\ \ref{sec:enhanced_spont_em} and \ref{sec:MWA_fermions} broaden the above discussion. In
Sect.\ \ref{sec:enhanced_spont_em} we show that there is a major difference how the condensate
affects light scattering and spontaneous emission, i.e.,\ spontaneous emission can probe
properties of the condensate beyond the structure factor. Eq.\ \ref{eq:net-rate} seems to imply
that enhancement of a process requires macroscopic population of a quantum state, i.e.,\ bosonic
quantum degeneracy. The discussion on matter wave amplification for fermions in Sect.\
\ref{sec:MWA_fermions} shows that this is not the case, and that collective enhancement is
possible even for fermions when they are prepared in a cooperative state.

\section{Determination of the static structure factor by Bragg spectroscopy}
\label{sec:Bragg_spect}

The matrix element or the structure factor in Eqs.\ \ref{eq:net-rate} and \ref{eq:anti-stokes-rate}
can be directly determined experimentally by light scattering.  The density of states (the $\delta$
function in Eq.\ \ref{eq:net-rate}) does not restrict the scattering since the photon energy is
much higher than the quasi-particle energy.  As a result, photons can be scattered into the full
solid angle and provide the necessary recoil energy of the atom by a small change ($\approx
10^{-9}$) in the frequency of the scattered photon.

\begin{figure}
    \begin{center}
    \includegraphics[height=55 mm]{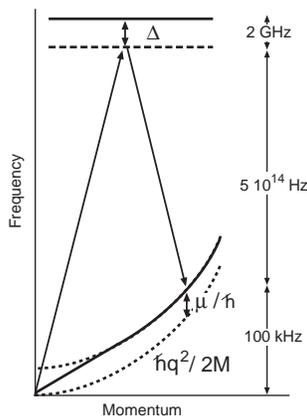}
    \caption{Energy level diagram for Bragg spectroscopy. The solid line represents the energy
    of an atom at momentum $\hbar q$, which is given by the Bogoliubov dispersion relation for a
    homogeneous weakly-interacting Bose-Einstein condensate (Eq.\ \ref{eq:Bogoliubov-energies}).
    Excitations can be created optically by stimulated light scattering using two laser beams which
    are both far detuned (about 2 GHz) from the atomic resonance. Momentum and energy are provided
    by absorption of one photon to a virtual excited level, followed by stimulated emission of a
    second, lower energy photon. For small momenta $\hbar q$, such that $\hbar q \ll M c_s$ the
    dispersion relation is phonon-like (linear). Here $M$ is the mass, and $c_s$ is the speed of
    sound which is related to the interaction energy $\mu$ by $\mu=M c_s^2$. For large momenta
    $\hbar q \gg M c_s$ the dispersion relation is particle-like (quadratic) offset from the energy
    of a free-particle excitation by a mean-field shift of $\mu$ ($\approx$ 5 kHz for our
    experiments).
     \label{fig:bogoliubov}}
     \end{center}
\end{figure}

\begin{figure}
    \begin{center}
    \includegraphics[height=2.7 in]{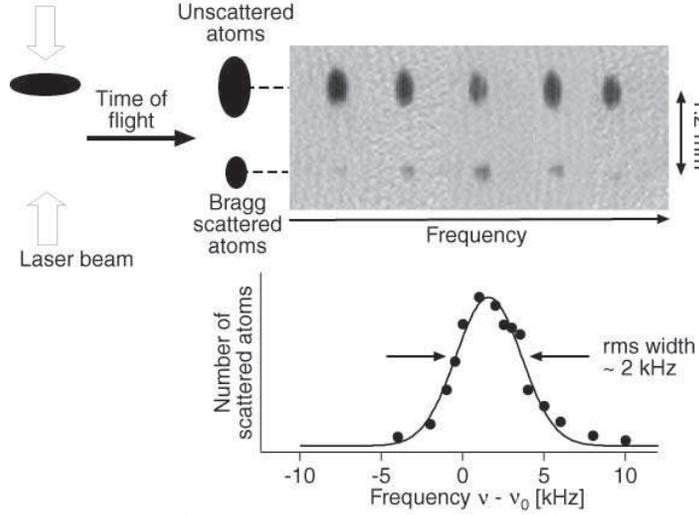}
    \caption{Bragg spectroscopy. Atoms were stimulated by two counter--propagating laser beams to
    absorb a photon from one beam and emit it into the other beam, resulting in momentum transfer
    to the atoms, as observed in ballistic expansion after 20 ms time of flight.  The number of
    scattered atoms showed a narrow resonance when the difference frequency $\nu$ between the two
    laser beams was varied around the recoil frequency $\nu_0$ of the atoms.
    \label{fig:bragg}}
    \end{center}
\end{figure}

When the light scattering is stimulated by a second laser beam, the momentum transfer $\hbar q$ is
determined by the angle $\theta$ between the two laser beams: $\hbar q = 2 \hbar k \sin(\theta /
2)$.  By varying the angle, one can probe both the phonon and free particle regime of the
Bogoliubov quasiparticles (Fig.\ \ref{fig:bogoliubov}). Fig.\ \ref{fig:bragg} illustrates our
experimental method for probing the response of a condensate at large momentum transfer.
Counter--propagating laser beams were incident perpendicular to the long axis of the cigar--shaped
condensate, which contained several million sodium atoms in the $F=1, m_F=-1$
state~\cite{mewe96bec}. After the momentum transfer, the condensate was allowed to freely expand,
allowing scattered atoms to separate spatially from the unscattered atoms, as shown in the figure.

For strong laser pulses and short times, the atoms undergo Rabi oscillations between the initial
state and the recoil state (see Sec.\ \ref{sec:Amplification of light in a dressed condensate}).
For longer times, we can use  rate equations (Eqs.\ \ref{eq:net-rate} and
\ref{eq:anti-stokes-rate}), and obtain the rate $W$ of transferring photons from one beam to the
other as
\begin{equation}
    W/N_0=(W_+ + W_-)/N_0= 2 \pi (\Omega_R/2)^2 S(q) [\delta(\omega -\omega_q^B)-\delta(\omega
    +\omega_q^B)].
    \label{eq:total-stim-rate}
\end{equation}
The two laser beams have wavevectors $\bf k$ and $\bf k-q$ and a difference frequency $\omega$.
The two-photon Rabi frequency $\Omega_R$ is given by $(\hbar \Omega_R/2)^2= |C|^2 n_k n_{k-q}$ (see
Sec.\ \ref{sec:four_wave_mixing}).

\begin{figure}
    \begin{center}
    \includegraphics[height=1.5 in]{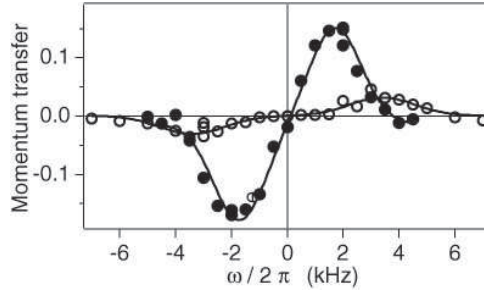}
        \caption{ Probing $S(q)$ by Bragg spectroscopy. Momentum transfer per
        particle, in units of $\hbar q$, is shown vs.\ the frequency difference $\omega / 2 \pi$
        between the two Bragg beams.  The two spectra were taken at different atomic densities.  The
        open symbols represent the phonon excitation spectrum for a trapped condensate (at a chemical
        potential $\mu / h = 9.2$ kHz, much larger than the free recoil shift of $\approx 1.4$ kHz).
        Closed symbols show the free-particle response of a twenty-three times more dilute
        (ballistically expanded) cloud. Lines are fits to the difference of two Gaussian line shapes
        representing excitation in the forward and backward directions. See Refs.\
        \protect\cite{stam99phon,stamp00leshouches} for more details. Figure is taken from Ref.\
        \protect\cite{stam99phon}. }
    \label{fig:phononspectra}
    \end{center}
\end{figure}

When $\omega$ is scanned the ``spectrum'' of a condensate consists of two peaks at $\pm
\omega_q^B$ (Fig. \ref{fig:phononspectra}). The strength (or integral) of each peak corresponds to
$S(q)$.  We refer to this method as Bragg spectroscopy since the basic process is Bragg scattering
of atoms from an optical standing wave.  A  full account of this method and the underlying theory
was given in our Les Houches notes \cite{stamp00leshouches} (see also
\cite{stam99phon,sten99brag,zamb00}).

We observed that the scattering rate was strongly suppressed when the momentum transfer $\hbar q$
became smaller than the speed of sound $c_s$ in the condensate (times the atomic mass $M$), i.e.,\
when the light scattering excited a phonon and not a free particle.  These observations are in
agreement with the Bogoliubov theory, which obtains the elementary excitations or Bogoliubov
quasiparticles as eigenstates of the Hamiltonian
\begin{equation}
    \label{eq:hamil_indist}
    {\cal H} =
    \sum_{k} \frac{\hbar^2 k^2}{2M} \hat{a}^\dagger_k \hat{a}_k
    + C \sum_{k,l,m,n} \hat{a}^\dagger_{l} \hat{a}^\dagger_n \hat{a}_k \hat{a}_m \delta_{l+n-k-m}.
\end{equation}
The constant $C$ can be expressed by the scattering length $a$ as
\begin{equation}
    \label{eq:C-indist}
    C=\frac{2\pi \hbar^{2} a}{M V}
\end{equation}
where $V$ is the condensate volume. The structure factor $S(q)$
(Eq.\ \ref{eq:def_s_q}) is the norm of the state vector
$\hat{\rho}({\bf q})^{\dagger}|g\rangle/\sqrt{N}$ with the atomic
density operator $\hat{\rho} ({\bf q})^{\dagger} = \sum_m
\hat{a}^\dagger_{m+q} \hat{a}_m$.  Only terms involving the
zero-momentum state $m=0$ yield significant contributions. Thus
$S(q)$ is the norm of the vector
\begin{equation}
    |e\rangle  \approx  \frac{(\hat{a}^\dagger_q \hat{a}_0 + \hat{a}^\dagger_0
    \hat{a}_{-q}) |g \rangle}{\sqrt{N}} \approx  (\hat{a}^\dagger_q  + \hat{a}_{-q}) |g \rangle
     =  |e^+\rangle + |e^-\rangle\,
    \label{state_vector_e}
\end{equation}
where we have replaced $\hat{a}^\dagger_0$ and $\hat{a}_0$ by $\sqrt{N}$ following the usual
Bogoliubov formalism \cite{bogo47}.

To calculate the norm of $|e \rangle$ explicitly, we transform to Bogoliubov operators
$\hat{b}^\dagger_{q} = u_q \hat{a}^{\dagger}_q + v_q \hat{a}_{-q}$, where $u_{q} = \cosh \phi_{q}$,
$v_{q} = \sinh \phi_{q}$ and $\tanh 2 \phi_{q} = \mu / (\hbar \omega_{q}^{0} + \mu)$ where $\hbar
\omega^{0}_{q} = \hbar^2 q^2 / 2 M$ is the free-particle kinetic energy and $\mu$ is condensate's
chemical potential.  This yields $S(q)=(u_q - v_q)^2$.   The static structure factor tends to $S(q)
\rightarrow \hbar q / 2 m c_s$ and vanishes in the long wavelength limit, as required of a
zero--temperature system with finite compressibility \cite{pric54}.  The speed of sound $c_s$ is
related to the chemical potential $\mu$ by $\mu=M c_s^2$. The resonances in the Bragg spectrum
occur at the energies $\hbar \omega_q^{B}$ of quasi-particles created by $\hat{b}^{\dagger}_{q}$
\begin{equation}
    \hbar \omega_q^{B} = \sqrt{ \hbar \omega^{0}_{q} (\hbar \omega^{0}_{q} + 2 \mu)
    }.
    \label{eq:Bogoliubov-energies}
\end{equation}

In our experiments we have observed the suppression of Bragg scattering at small angles.  This
implies an overall suppression of Rayleigh scattering which will be more pronounced when the speed
of sound becomes comparable or larger than the recoil velocity.  By integrating the static
structure factor $S(q)$ over all possible scattering angles and accounting for the dipolar
emission pattern, we find that Rayleigh scattering from a BEC is suppressed by a factor
\cite{stam99phon}
\begin{eqnarray}
    \label{equ:f_bose_scatt}
    &&F_{\text{Bose}}^{\text{scatt}} = \frac{k_s}{\sqrt{k_s^2+k_L^2}}
    \left(\frac{15}{8} \,\frac{k_s^5}{k_L^5} \, +  \, \frac{23}{8} \, \frac{k_s^3}{k_L^3} \, + \, 2 \,
    \frac{k_s}{k_L} \, + \, \frac{k_L}{k_s} \right)
    \\ \nonumber && \ \ - \, \left(\frac{15}{8} \, \frac{k_s^6}{k_L^6} \, + \,
    \frac{9}{4} \, \frac{k_s^4}{k_L^4} \, + \, \frac{3}{2} \, \frac{k_s^2}{k_L^2} \right) \,
    \tanh^{-1} \left( \frac{k_L}{\sqrt{k_s^2+k_L^2}} \right) \ .
    \\ \nonumber
\end{eqnarray}
where $\hbar k_L$ is the incident photon momentum and $\hbar k_s=M
c_s$ is the momentum of an atom moving at the speed of sound.

The long-wavelength suppression of $S(q)$ reveals a remarkable
many-body effect.  For free particles, the matrix element for
momentum transfer is always 1, which reflects the fact that the
operator $\exp({\rm i}{\bf q \cdot r})$ connects an initial state
with momentum ${\bf p}$ to a state with momentum ${\bf p}+\hbar
{\bf q}$ with unity overlap.  For an interacting Bose-Einstein
condensate, this overlap vanishes in the long-wavelength limit.
As we have discussed in previous publications
\cite{stam99phon,stamp00leshouches}, this suppression of momentum
transfer is due to an destructive interference between the two
pathways for a condensate to absorb momentum $\hbar \bf q$ and
create a quasi-particle: one pathway annihilates an admixture with
momentum $- \hbar {\bf q}$, the other creates an admixture at
momentum $+ \hbar {\bf q}$.

The leading terms in stimulated light scattering do not depend on temperature.  The rates $W_+$ and
$W_-$ (Eqs. \ref{eq:net-rate} and \ref{eq:anti-stokes-rate}) are independent of the thermally
excited population of quasiparticles $N_q$ and $N_{-q}$ in the limit of large $n_k,\:n_{k-q}\gg
N_{\pm q}$, i.e.,\ when the scattering is stimulated by a second laser beam.  For spontaneous
scattering ($n_{k-q}=0$), one obtains for the total scattering rate $W$ instead of Eq.
\ref{eq:total-stim-rate}
\begin{equation}
    W/N_0=  \frac{2 \pi}{\hbar^2} |C|^2 S(q) n_k \left[(1+N_q) (\delta(\omega - \omega_q^B)+N_{-q}
    \delta(\omega +\omega_q^B)) \right].
    \label{eq:total-spont-rate}
\end{equation}
Absorption of and bosonic stimulation by thermally excited quasi-particles become important when
the temperature is comparable to or larger than the quasi-particle energy $\hbar \omega_q^B$. Eq.\
\ref{eq:total-spont-rate} is proportional to the temperature dependent dynamic structure factor
$S_T({\bf q},\omega)$\cite{grif93}.

\section{Atomic collisions in a Bose-Einstein condensate}
\label{sec:BEC-collisions}

Before we discuss the analogies and differences between the scattering of light and massive
particles, we make general remarks on particle scattering. We raise the question under what
conditions can two matter waves penetrate each other without scattering. This includes the question
what happens when two atom laser beams cross each other.

As we have seen in Sect.~\ref{sec:scattering}, the scattering rate is proportional to the structure
factor $S(\bf q)$, which is a measure of the density-density correlations at wavevector ${\bf q}$
associated with the momentum transfer. It consists of two parts, one reflects the average atomic
density $\tilde{\rho}({\bf q})= \langle g | \hat{\rho}({\bf q})| g \rangle$ and the other one the
fluctuations $\delta\hat{\rho} ({\bf q})=\hat{\rho}({\bf q})-\tilde{\rho}({\bf q})$ (see also
Eq.~\ref{eq:sqw}).
\begin{eqnarray}
S({\bf q})&=& \frac{1}{N}\langle g |
    (\tilde{\rho}({\bf q})+\delta\hat{\rho}({\bf q}))
    (\tilde{\rho}^*({\bf q})+\delta\hat{\rho}^\dagger({\bf q}))
    | g \rangle \\
    &=& \frac{1}{N}\left(
    \tilde{\rho}({\bf q})\tilde{\rho}^*({\bf q})
    +\langle g |\delta\hat{\rho}({\bf q})\delta\hat{\rho}^\dagger({\bf q})| g \rangle
    \right)
\end{eqnarray}

In a system without fluctuations ($\delta\hat{\rho}({\bf q})=0$), there is only scattering when the
stationary density modulation allows for a momentum transfer at wavevector ${\bf q}$ (
$\tilde{\rho}({\bf q})\neq 0$) and the incident particles fulfil the Bragg condition. One well
known example is X-ray scattering off a crystal lattice.  Electrons in such a lattice form
stationary Bloch waves (superpositions of plane waves and Bragg scattered plane waves) and
propagate without attenuation. Scattering only occurs at irregularities of the lattice or thermal
fluctuations.

A Bose-Einstein condensate might appear perfectly ordered. However, as we have seen in
Sect.~\ref{sec:Bragg_spect}, it has density fluctuation similar to the classical ideal gas. The
structure factor only differs from an ideal gas when the momentum transfer is comparable or less
than the speed of sound (times the mass $M$). In the case of electromagnetic waves, these density
fluctuation cause Rayleigh scattering. In close analogy, if a matter wave propagates through a
condensate, there will be elastic scattering. On length scales larger than the healing length, the
condensate has reduced density fluctuations, and we have seen that electromagnetic radiation of
sufficiently long wavelength can propagate with only little scattering
(Eq.~\ref{equ:f_bose_scatt}). In the next section, we will find the equivalent result for long
wavelength matter waves, but also a new effect, namely the complete suppression of scattering for
velocities below the Landau critical velocity.

\begin{figure}
    \epsfxsize=70mm \centerline{\epsfbox{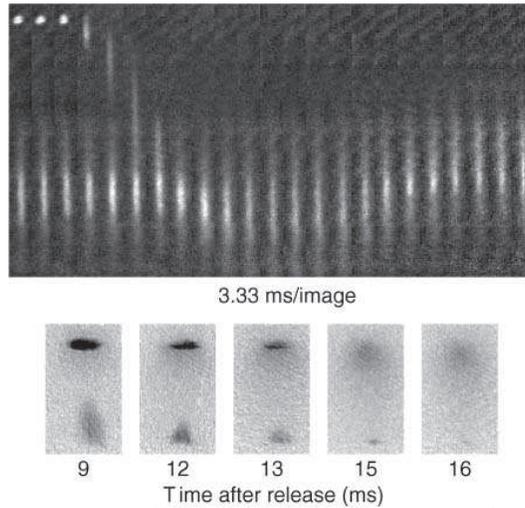}}
        \caption{Collisions between two condensates.  One condensate was held up in the magnetic trapping
        potential by a blue-detuned light sheet and suddenly released.  A series of phase-contrast images
        (upper part) and time-of-flight absorption images display the collision.  The phase contrast
        pictures show the spreading of the upper condensate during acceleration, its collision, and
        merging with the stationary cloud. The time-of-flight pictures represent the velocity
        distribution. Before the scattering event, there were two distinct peaks representing the
        accelerated and the stationary condensate.  After the collision, only one cloud was left which was
        heated up by the collision.}
    \label{fig:Magnetic_atom_accelerator}
\end{figure}

The scattering between two Bose-Einstein condensates can be nicely demonstrated by creating two
condensates in a magnetic trap, one sitting at the bottom, the other one held up in the trapping
potential by a blue-detuned light sheet which forms a repulsive potential for atoms. This initial
situation was created by splitting a condensate into two halves using a light sheet
\cite{andr97int} and then shifting the center of the magnetic trap by applying a magnetic field
gradient.  When the light sheet was switched off, one condensate accelerated and slammed into the
other.  The violent collision was observed {\it in situ} by phase contrast imaging and also
analyzed by absorption imaging after ballistic expansion (Fig.\
\ref{fig:Magnetic_atom_accelerator}).

\begin{figure}
    \epsfxsize=30mm \centerline{\epsfbox{s_wave_ring.eps}}
        \caption{
        Observation of elastic collisions between the condensate (lower black dot) and Bragg
        diffracted atoms (upper spot).  The image is taken after a time-of-flight of 30 ms, and shows the
        velocity distribution after the collision.  The products of the collision are distributed over a
        sphere in momentum space leading to the observed $s$-wave halo. The height of the image is 3.2 mm.
        }
    \label{fig:s_wave_ring}
\end{figure}

The method shown in Fig.\ \ref{fig:Magnetic_atom_accelerator} suffers from a large velocity spread
of the incident condensate.  A narrower velocity distribution can be achieved by Bragg scattering.
With counter-propagating laser beams, we transferred two photon recoil momenta to some of the
atoms.  After the recoiling atoms had traversed the condensate the velocity distribution was
probed in time-of-flight imaging  and showed the characteristic $s$-wave halo (Fig.\
\ref{fig:s_wave_ring}).

Such collisions between indistinguishable atoms of mass $M$ are
described by the interaction Hamiltonian in Eq.\
\ref{eq:hamil_indist}. This Hamiltonian is not only responsible
for the interactions between two condensates, but also for the
mean-field energy of a single condensate ($C \hat{a}^\dagger_{0}
\hat{a}^\dagger_0 \hat{a}_0 \hat{a}_0$), pair correlations and
quantum depletion ($C \sum_{q} \hat{a}^\dagger_{q}
\hat{a}^\dagger_{-q} \hat{a}_0 \hat{a}_0$) and the damping of
collective excitations where a quasi-particle in mode $k$ decays
into two other excitations (Beliaev damping, $2C \sum_{l,n}
\hat{a}^\dagger_{n} \hat{a}^\dagger_{l} \hat{a}_k \hat{a}_0
\delta_{l+n-k-0}$) or promotes an existing quasi-particle in mode
$l$ to higher energy (Landau damping, $4C \sum_{k,n}
\hat{a}^\dagger_{0} \hat{a}^\dagger_{n} \hat{a}_l \hat{a}_k
\delta_{0+n-l-k}$).

Several papers discussed the outcoupling of atoms in mode $k$ from
a condensate at rest (mode 0) and only considered the term $2C
\hat{a}^\dagger_{k} \hat{a}^\dagger_0 \hat{a}_0 \hat{a}_k$ which
represents the mean-field repulsion between the two matter waves
(see, for example, \cite{ball97,band99atom}). This term leads to
a distortion of the outcoupled beam. However, this two-mode
approximation neglects the scattering into the empty modes,
described by
\begin{equation}
    \label{eq:swavehamil}
    {\cal H^\prime} = 2C \sum_{l,n} \hat{a}^\dagger_{l}
    \hat{a}^\dagger_n \hat{a}_{k} \hat{a}_0 \delta_{l+n-k-0}
\end{equation}
In the following we want to show that this interaction leads to a
scattering rate for particles in mode $k$
\begin{equation}
    W_{\rm tot}=n_0 \sigma v
    \label{eq:coll_cross_sect}
\end{equation}
with the condensate density $n_0=N_0/V$, the cross-section $\sigma=8 \pi a^2$ and the relative
velocity $v=\hbar k/M$. Eq.\ \ref{eq:coll_cross_sect} therefore describes not only classical
particles which collide with a cross-section $\sigma$, but also the quantum mechanical interaction
between two matter waves.

We assume that the incident velocity is much larger than the speed of sound $v \gg c_s$.
Therefore, there is no distinction between free particles and quasi-particles.  In the
center-of-mass frame the relevant terms in the Hamiltonian (Eq.\ \ref{eq:swavehamil}) are
\begin{equation}
    \label{eq:swavehamil2}
    {\cal H^{\prime}} = 4 C \sum_{q}{}^{\prime}\; \hat{a}^\dagger_{q}
    \hat{a}^\dagger_{-q} \hat{a}_{k/2} \hat{a}_{-k/2},
\end{equation}
where the prime indicates summation over one hemisphere. The additional factor of 2 (compared to
Eq.\ \ref{eq:swavehamil}) accounts for the fact that the final state $({\bf q,-q})$ appears twice
in the summation (as ${\bf q,-q}$ and ${\bf -q,q}$), we only count the number of states in one
hemisphere). Considering the collision of one particle with momentum $\hbar {\bf k}$ with a pure
condensate, we write the initial and final states $|N_{k/2}, N_{q} ; N_{-k/2}, N_{-q}\rangle$ (see
Eq.\ \ref{eq:states}) as
\begin{eqnarray}
    \label{eq:s-wavestates_p}
    |i\rangle &=&  |1, 0 ; N_0, 0\rangle \nonumber  \\
    |f\rangle &=&  |0, 1 ; N_0-1,1\rangle.
\end{eqnarray}

The square of the matrix element for the Hamiltonian (Eq.\ \ref{eq:swavehamil2}) between those two
states is $|M|^{2}=16|C|^{2}N_{0}$.  From Eq.\ \ref{eq:rate} (Fermi's golden rule) the scattering
rate follows as
\begin{eqnarray}
    \label{eq:swaverate}
    W_{\rm tot} &=& \frac{2 \pi}{\hbar} \sum_{f} |M|^2  \delta(E_{f} - E_{i}) \nonumber \\
    &=& \frac{2 \pi}{\hbar} |M|^{2} \int \!\! \left(\frac {{\rm d} \rho}{{\rm d} \!E_{f}} \right)
    \delta(E_{f} - E_{i}) \; {\rm d}\!E_{f}
\end{eqnarray}
with the final and initial energies $E_{f}=2 \times {\hbar^{2}q^{2}}/{2M}$, $E_{i}=2 \times
{\hbar^{2}(k/2)^{2}}/{2M}$.  The density of final states ${{\rm d}\!\rho}/{\rm d} \!E_{f}$ is equal
to half the density of single particle states evaluated at $E_{\rm single}=E_f/2$. The density of
single particle states at energy $E$ is
\begin{equation}
    \frac{{\rm d}\rho}{{\rm d}E}=\frac{1}{2} \times \frac{V}{(2\pi)^{3}} 2\pi
    \left(\frac{2M}{\hbar^{2}}\right)^{3/2} E^{1/2}
\end{equation}
The additional factor of $(1/2)$ comes from the fact that for
pairs of indistinguishable particles, we only count the number of
states in one hemisphere.

Thus, the density of final states is
\begin{equation}
    \label{eq:density_finalstates}
    \frac{{\rm d}\rho}{{\rm d}E_{f}} = \frac{1}{2}
    \left.\frac{{\rm d}\rho}{{\rm d}E} \right|_{E=E_{f}/2}
    = \frac{V}{(2\pi)^{3}} \frac{\pi}{2}
    \left(\frac{2M}{\hbar^{2}}\right)^{3/2} (E_{f}/2)^{1/2}
\end{equation}
Substituting Eq.\ \ref{eq:density_finalstates} into Eq.\ \ref{eq:swaverate} gives $q=k/2$ and
finally Eq.\ \ref{eq:coll_cross_sect}.

In the center-of-mass frame, the scattered particles occupy a shell in momentum space at $q=k/2$
(see Fig.\ \ref{fig:s_wave_ring}).  The quantum-mechanical origin of the relative velocity $v$ in
Eq.\ \ref{eq:coll_cross_sect} is the density of final states which is proportional to
$\sqrt{E}\propto v$. The scattering of atoms into empty modes is not described by the
Gross-Pitaevskii equation (which only describes macroscopically occupied modes), but can be
accounted for by introducing a complex scattering length into the Gross-Pitaevskii equation
\cite{band00loss}.

For the scattering of impurity atoms with the same mass as the condensate atoms the constant $C$
in Eq.\ \ref{eq:hamil} is
\begin{equation}
    \label{eq:C-dist}
    C=\frac{4\pi \hbar^{2} a}{M V},
\end{equation}
or more generally $C=2 \pi \hbar^2 a/ \mu V$ where $\mu$ is the reduced mass and $a$ is now the
scattering length for collisions between condensate and impurity atoms. The factor of two
difference between Eqs.\ \ref{eq:C-dist} and \ref{eq:C-indist} is necessary to avoid double
counting of identical atom pairs.  If we repeat the above derivation for impurity atoms we obtain
a cross-section of $\sigma=4 \pi a^2$, whereas the cross-section for indistinguishable particles is
twice as large.  This reflects several factors of 2 and 4:  The factor of $(1/2)$ in the constant
$C$; the factor of 4 in Eq.\ \ref{eq:swavehamil2} which expresses that each initial and final
state appears four times in Eq.\ \ref{eq:hamil_indist}; the momentum integral for
indistinguishable particles extends only over the hemisphere whereas distinguishable particles
have twice the number of possible final states.

We used images like Fig.\ \ref{fig:s_wave_ring} (for indistinguishable atoms) and Fig.\
\ref{collisions}a for impurity atoms (see following section) to determine the collision cross
sections. The cross-section for indistinguishable atoms was found to be $2.1 \pm 0.3$ times larger
than the one for impurities in agreement with the expected factor of two \cite{chik00}.

\section{Suppression of impurity collisions}
\label{sec:collisions}

The key difference between the scattering of light and massive particles (or impurities) is their
energy-momentum dispersion relation. The dispersion relation for impurities is $E_{k}=(\hbar
k)^{2}/2M$, whereas for light, $E_{k}=\hbar k c$ with $c$ denoting the speed of light.
This difference is responsible for the complete suppression of impurity scattering at low
velocities.  For simplicity, we assume equal mass $M$ for the impurity and condensate atoms.

Energy-momentum conservation, the $\delta$ function in Eq.\ \ref{eq:net-rate}, requires $E_k -
E_{k-q} = \hbar \omega_q^B$. For impurity particles, the l.h.s.\ is always less than $v \hbar q$,
where $v=\hbar k/M$ is the initial velocity of the impurities. Thus, collisions with the
condensate are only possible, if this maximum energy transfer is sufficient to excite a
quasi-particle, i.e.,\ $v>\min(\omega_q^B/q)=v_{L}$, where $v_{L}$ is the Landau critical
velocity~\cite{land41} for superfluidity below which no dissipation occurs because the density of
final states vanishes.

For the excitation spectrum of the condensate (Eq.\ \ref{eq:Bogoliubov-energies}), the Landau
velocity is the Bogoliubov speed of sound $v_{L}=c_s=\sqrt{\mu/M}$.  Impurity particles moving below
this speed cannot dissipate energy in collisions.  In contrast, for photons, ${\rm d}E_k/{\rm
d}(\hbar k)=c \gg c_s$, and scattering is always possible.

In the perturbative limit (no stimulation by the final
occupation), the total rate of scattering $W_{\rm tot}$ is given
by integrating Eq.\ \ref{eq:net-rate} over all possible momentum
transfers.  In the derivation above for large incident velocities
we absorbed the angular integral in the density of states since
the scattering in the center-of-mass frame is isotropic.  Now it
is necessary to consider the scattering angles explicitly.
Furthermore, we stay in the frame where the condensate is at
rest. Straightforward transformations lead to
\begin{eqnarray}
        W_{\rm tot}
        & = & \frac{2\pi |C|^{2}}{\hbar^{2}} N_{0}
        \sum\nolimits_{\bf q} {S(q)} \,\,
        \delta \left(\frac{\hbar {\bf k \cdot
        q}}{M}-\frac{\hbar q^{2}}{2 M}- \omega_{q}^{B} \right) \nonumber \\
        &=&  (N_{0}/V)\left(\frac{2\hbar a}{M}\right)^{2}
        \int \!\! {\rm d}q {\rm d}\Omega \,\, q^{2} S(q) \,\,
        \delta \left(\frac{\hbar  k q \cos\theta}{M}-\frac{\hbar q^{2}}{2 M}-
        \omega_{q}^{B} \right) \nonumber \\
        &=&  2\pi (N_{0}/V)\left(\frac{2\hbar a}{M}\right)^{2} \frac{1}{v}
        \int^{Q}_{0} \!\! {\rm d}q \,\, q S(q) \nonumber \\
        &=& (N_{0}/V) \: \sigma(\eta) \: v,
    \label{eq:total_rate}
\end{eqnarray}
where $\hbar Q= M v (1-1/\eta^{2})$ is the maximum possible momentum transfer, and $\eta=v/c_s$
must be larger than 1.  The collision cross-section is $\sigma(\eta) = \sigma_0 F(\eta)$ where
$\sigma_0 = 4 \pi a^2$. For $\eta<1$, $F(\eta)=0$ and for $\eta>1$,
$F(\eta)=1-1/\eta^{4}-\log(\eta^{4})/\eta^{2}$.

\begin{figure}
    \epsfxsize=50mm \centerline{\epsfbox{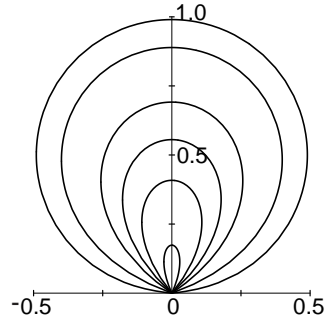}}
        \caption{Momentum transfer in
        collisions. The polar diagram shows the momentum transfer (in units of the initial momentum) vs.\
        scattering angle $\theta$ for different values of $\eta=v/c_s$ (10, 3, 1.8, 1.5, 1.3, 1.1). As the impurity
        velocity approaches the speed of sound ($\eta \rightarrow 1$), the scattering cone shrinks to zero
        solid angle.}
    \label{phase_space}
\end{figure}

\begin{figure}
    \epsfxsize=70mm
    \centerline{\epsfbox{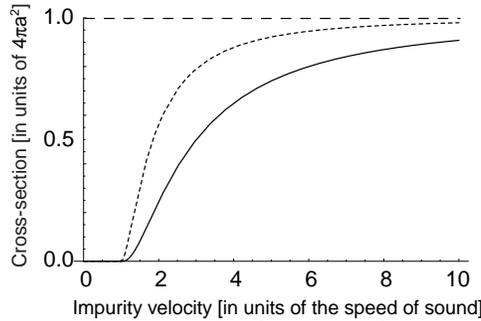}}
        \caption{Suppression of collisions. Shown is the
        suppression factor as a function of the impurity velocity
        (normalized by the speed of sound $c_s$, solid line). The dotted
        line represents the suppression due to phase-space restriction
        alone (i.e.,\ setting the structure factor $S(q)=1$).}
    \label{coll_fc_theory}
\end{figure}

The suppression factor $F(\eta)$ is determined by two factors:
the phase space restriction due to the $\delta$ function in
Eq.~\ref{eq:total_rate}, and additional suppression at low
momentum transfers by the structure factor of the condensate. For
decreasing velocity, the possible scattering angles $\theta$
become restricted to a forward scattering cone
($\theta<\arccos(1/\eta)$), which shrinks to zero solid angle at
the Landau critical velocity (Fig.\ \ref{phase_space}).  This
reflects that near the Landau velocity, the scattered particle
has ``difficulties'' to provide enough energy per momentum to
create phonons.  The maximum energy transfer occurs when the
momentum transfer is collinear with the incident velocity, i.e.,\
for small angle scattering angles. A graph of the suppression
factor as function of impurity velocity is shown in Fig.\
\ref{coll_fc_theory}.

Experimentally, the Landau critical velocity can usually only be observed by moving
\emph{microscopic} particles through the superfluid which do not create a macroscopic flow
pattern.  Studies of superfluidity with microscopic objects were pursued in liquid $^{4}$He by
dragging negative ions through pressurized $^{4}$He~\cite{meye61,allu77}, and by scattering
$^{3}$He atoms off superfluid $^{4}$He droplets~\cite{harm99}.

\begin{figure}
    \epsfxsize=90mm
    \centerline{\epsfbox{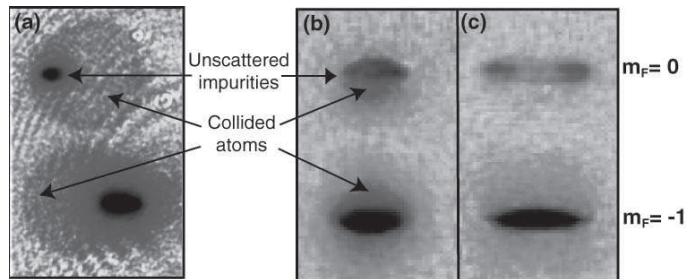}}
        \caption{ Observation of elastic collisions between the condensate and impurity atoms. (a)
        Impurities traveling at 6 cm/s along the radial axis (to the left in images) were scattered into
        an $s$-wave halo. Absorption image after 50 ms of time-of-flight
        shows the velocity distribution
        after collisions between the condensate (bottom) and the outcoupled $m_{F}=0$ atoms (top),
        spatially separated by a Stern-Gerlach type magnetic field gradient. The collisional products are
        distributed over a sphere in momentum space. The image is 4.5 $\times$ 7.2 mm. (b) Similar image
        as (a) shows the collisional products (arrow) for impurity atoms (top) traveling at 7 mm/s along
        the condensate axis (upward in image). For this image, $v_{g}/c_s =2.7$ (see text). Collisions are
        visible below the unscattered impurities. (c) Similar image as (b) with $v_{g}/c_s =1.6$. Collisions
        are suppressed. The momentum distribution of the outcoupled atoms was distorted by mean-field repulsion.
        The images are 2.0 $\times$ 4.0~mm. Figure is taken from Ref.\ \protect\cite{chik00}.}
    \label{collisions}
\end{figure}

To study the effects of impurities interacting with the condensate, we created microscopic
impurity atoms using a stimulated Raman process which transferred a small fraction of the
condensate atoms in the $|F=1,m_{F}=-1\rangle$ hyperfine state into an untrapped hyperfine state
$|F=1,m_{F}=0\rangle$ with a well-defined initial velocity~\cite{hagl99}. The initial velocity
could be adjusted between zero and two single-photon recoil velocities by varying the angle
between the two Raman beams.  As these impurities traversed the condensate, they collided with the
stationary condensate, resulting in a redistribution of the impurity momenta which was detected by
a time-of-flight analysis \cite{chik00} (Fig.\ \ref{collisions}).

\begin{figure}
    \epsfxsize=60mm
    \centerline{\epsfbox{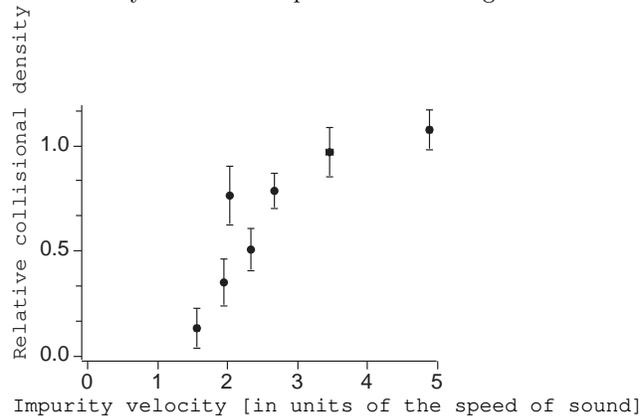}}
        \caption{Onset of superfluid suppression of collisions.  Shown is the observed collisional density
        normalized to the predicted one in the limit of high velocities as a function of
        $\overline{\eta}=v_{g}/c_s$, which is a measure of the impurity velocity in units of the
        condensate's speed of sound. The error bars represent the statistical uncertainty. Data is taken
        from Ref.\ \protect\cite{chik00}. }
    \label{coll_maindata}
\end{figure}

To probe for the suppression of collisions, the impurity velocity
has to be varied around the speed of sound. For that, we produced
impurity atoms at low velocities (7 mm/s) and varied the speed of
sound by changing the condensate density.  The small axial
velocity imparted by Raman scattering allowed us to identify
products of elastic collisions in time-of-flight images (Fig.\
\ref{collisions}b, c) since collisions with the stationary
condensate redistributed the impurity atoms toward lower axial
velocities. However, the impurity velocity was predominantly
determined by the gravitational acceleration $g$, which imparted
an average velocity of $v_{g}=\sqrt{g l_{z}}$ where $l_{z}$ is the
Thomas-Fermi diameter of the condensate in the $z$-direction.
Thus, the effect of superfluidity on impurity scattering depends
primarily on the parameter $\overline{\eta} = v_g / c_s$ which is
the ratio of the typical impurity velocity $v_g$ to the speed of
sound $c_s$ at the center of the condensate.

A time-of-flight analysis of impurity scattering for the case of a low-density condensate (small
$c_s$) and large condensate radius (large $v_g$) is shown in Fig.\ \ref{collisions}b. The effect of
collisions is clearly visible with about 20\% of the atoms scattered to lower axial velocities
(below the unscattered impurities in the image). In contrast, in the case of tight confinement,
the condensate density is higher (larger $c_s$) and its radius is smaller (smaller $v_g$), and the
collision probability is greatly suppressed due to superfluidity (Figs.\ \ref{collisions}c and
\ref{coll_maindata}).

\begin{figure}
    \epsfxsize=60mm
    \centerline{\epsfbox{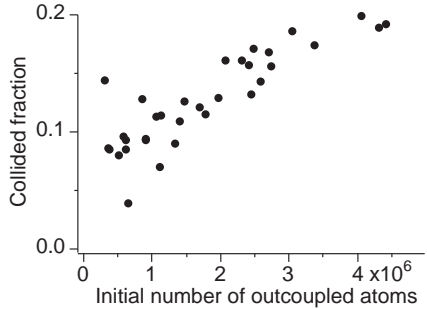}}
    \caption{ Collective amplified elastic scattering in a Bose-Einstein condensate. Shown is the
    fraction of collided atoms vs.\ the number of outcoupled atoms. For this data, $v_{g}/c_s$=4.9 and
    the chemical potential was 1.8 kHz. Figure is taken from Ref.\ \protect\cite{chik00}. }
    \label{coll_enhance}
\end{figure}

The bosonic stimulation factor in Eq.\ \ref{eq:net-rate} becomes relevant if the final states are
populated, either by scattering or thermally.  We observed that the fraction of collided atoms
increased with the number of outcoupled impurities (Fig.~\ref{coll_enhance}). For a large
outcoupled fraction, population $n_{k-q}$ and $N_q$ is built up in the final states and stimulates
further scattering. This collisional amplification is not directional, and is similar to the
recently observed optical omnidirectional superfluoresence~\cite{lvov99}.

Gain of momentum and thus transfer of energy from the condensate to the impurity atoms is
impossible at zero temperature, but may happen at finite temperature due to the presence of thermal
excitations (the $N_{-q}$ term in Eq.\ \ref{eq:anti-stokes-rate})\footnote{We are grateful to S.
Stringari for pointing out the importance of finite-temperature effects.}. Thus finite temperature
enhances the elastic cross-section by two effects: Absorption of quasi-particles (anti-Stokes
process) and stimulation of momentum transfer by the final state population (Stokes process).

\begin{figure}
    \epsfxsize=70mm
    \centerline{\epsfbox{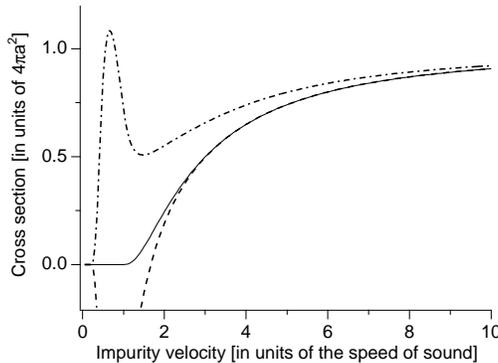}}
    \caption{Temperature dependent cross-section vs.\ impurity velocity. Shown is the
    cross-section at zero temperature (solid line) and at a finite temperature $kT=\mu$ which is
    typical for our experimental conditions (dash-dotted line). The finite temperature
    cross-section includes collisions involving thermally occupied quasi-particles where the
    impurities lose or gain energy. In the experiment, we measured the number of impurities which
    lost its energy minus the number which gained energy. Thus, the experimental measured
    cross-sections (Fig.\ \protect\ref{coll_maindata}) should be compared to $\sigma_{\rm
    coll,loss} -\sigma_{\rm coll,gain}$ (dashed line).}
    \label{finitetemp}
\end{figure}

Fig.\ \ref{finitetemp} shows the dramatic variation of the elastic scattering cross-section with
temperature. However, the finite temperature did not affect our data in a major way:  Due to
gravitational acceleration we couldn't probe the velocity regime well below the Landau critical
velocity where only thermally assisted collisions are possible.  Furthermore, when we counted the
number of collided atoms we had to use a background subtraction method where we subtracted the
small signal of the energy gain collisions from the energy loss collisions (see Ref.\
\cite{chik00} for details), thus cancelling most of the finite-temperature effects.

\section{Suppression of dissipation for a moving macroscopic object}
\label{sec:macroscopic_flow}

So far, we have discussed the suppression and enhancement of
microscopic processes (light scattering and impurity collisions).  The suppression of dissipation
is even more dramatic on the macroscopic scale.  The flow of liquid $^4$He and the motion of
macroscopic objects through it are frictionless below a critical velocity \cite{nozi90}.
Recently, we have explored such frictionless flow in a gaseous BEC \cite{rama99,onof00sup}.

The microscopic and macroscopic cases bear many parallels.  The onset of scattering or dissipation
has two requirements:  one needs final states which conserve energy and momentum, and an overlap
matrix element which populates these states.  In the case of macroscopic flow, the first
requirement leads to a critical velocity for vortex creation and the second requirement addresses
the nucleation process of vortices.

The Landau criterion for superfluidity shows that excitations with momentum $p$ and energy $E(p)$
are only possible when the relative velocity between the fluid and the walls or a macroscopic
object exceeds the Landau critical velocity $v_L$ which is given by $v_L = \min(E(p)/p)$ (see e.g.\
\cite{huan87,nozi90}). A similar criterion applied to vortex formation yields
\begin{equation}
    v_{c}  = \frac{E_{\rm vortex}}{I_{\rm vortex}}
    \sim \frac{\hbar}{{M D}}\ln \left( {\frac{D}{\xi }} \right)
    \label{eq:feyn}
\end{equation}
where $I_{\rm vortex} = \int{p \;{\rm d}^3\!r}$ is the integrated
momentum of the vortex ring or line pair, $E_{\rm vortex}$ is its
total energy, $D$ is the dimension of the container, and $\xi$
the core radius of a vortex which in the case of dilute gases is
the healing length $\xi = 1/\sqrt{8 \pi \rho a}$. Ref.\
\cite{nozi90} derived Eq.\ \ref{eq:feyn} for vortex rings with a
maximum radius $D$, Ref.\ \cite{cres00} looked at pairs of line
vortices at distance $D$. Feynman \cite{feyn55} found a similar
result for superflow through a channel of diameter $D$.

An analogous result is obtained for a Bose condensed system placed under uniform rotation with
angular velocity $\Omega$.  A vortex becomes energetically allowed when its energy $E'$ in the
rotating frame drops to zero,
\begin{equation}
    E' = E - \Omega L = 0,
\end{equation}
where $E$ and $L$ are the energy and angular momentum in the laboratory frame.  This defines a
critical {\em angular} velocity below which a vortex cannot be sustained due to conservation of
angular momentum and energy~\cite{lund97}:

\begin{equation}
    \Omega_c = \frac{E_{\rm vortex}}{L_{\rm vortex}} \sim
    \frac{\hbar}{M D^2}\ln{\left(\frac{D}{\xi}\right)}.
    \label{eq:vortex}
\end{equation}

The critical velocity at the wall of the rotating container,
$v_{c}=D \Omega_c$, agrees with Eq.\ \ref{eq:feyn}. However, Eqs.\
\ref{eq:feyn} and \ref{eq:vortex} only reflect the energy and
momentum required to generate vortices, and do not take into
account the nucleation process.  If the scattering particle is
macroscopic in size, the coupling is between the ground state and
a state containing a vortex. Populating such a state requires
nucleation of the vortex by the perturbing potential, which
usually does not occur until higher velocities are reached than
those predicted by Eqs.\ \ref{eq:feyn} and \ref{eq:vortex}. The
other option, the formation of the vortex by macroscopic quantum
tunneling between the two states is an extremely slow process. In
recent experiments in which a Bose condensate was placed in a
rotating potential, the critical angular velocity for the
formation of a single vortex was observed to be 1.6 times higher
than the value given by Eq.\ \ref{eq:vortex} \cite{madi00}.  This
discrepancy may be due to a nucleation barrier associated with
the excitation of surface modes, as some authors have recently
suggested \cite{fede99A,isos99vort}.

To study frictionless flow in a Bose-condensate, we focused an
argon ion laser beam (at 514 nm) onto the condensate, which
repelled atoms from the focus. The laser beam was scanned back
and forth along the axial direction of the condensate, creating a
moving ``hole'' that simulated a macroscopic object. Rapid
sequence phase-contrast imaging allowed us to directly measure
the density profile of the superfluid around the moving laser
beam.

For a weakly interacting Bose-condensed gas at density $\rho({\bf
r})$ and chemical potential $\mu({\bf r})$, pressure is identical
to the mean-field energy density $ P = \mu({\bf r}) \rho({\bf
r})/2$ \cite{dalf99rmp}. A drag force arises due to the pressure
difference across the moving object. The chemical potential is
given by $\mu({\bf r},t)=g \rho({\bf r},t)$, where $g = C V =4
\pi \hbar^2 a /M$ is the strength of two-body interactions. The
drag force $F$ is given by
\begin{equation} F \simeq  g S \rho \Delta \rho=S \mu \Delta \mu/g
\label{eq:drag_force}
\end{equation}
where $\Delta \rho$ and $\Delta \mu$ are the differences in density and chemical potential across
the stirring object, and $S$ the surface area the macroscopic object presents to the condensate.

\begin{figure}[htbf]
    \epsfxsize=70mm \centerline{\epsfbox{phase_contrast.eps}}\vspace{0.1cm}
    \caption{Pressure
    difference across a laser beam moving through a condensate. On the left side {\em in situ}
    phase contrast images of the condensate are shown, strobed at each stirring half period: beam
    at rest (top); beam moving to the left (middle) and to the right (bottom). The profiles on the
    right are horizontal cuts through the center of the images. The stirring velocity  and the
    maximum sound velocity were 3.0 mm/s and 6.5 mm/s respectively. Figure is taken from Ref.\
    \protect \cite{onof00sup}. }
    \label{fig:strobed}
\end{figure}
If the laser beam is stationary, or moves slowly enough to preserve the superfluid state of the
condensate, there will be no gradient in the chemical potential across the laser focus, and
therefore zero force according to Eq.\ \ref{eq:drag_force}. A drag force between the moving beam
and the condensate is indicated by an instantaneous density distribution $\rho(\bf{r},t)$ that is
distorted asymmetrically with respect to the laser beam. Fig.\ \ref{fig:strobed} shows phase
contrast images strobed at half the stirring period (where the laser beam is in the center of the
condensate).  A bow wave and stern wave form in front of and behind the moving laser beam,
respectively.  We define the asymmetry $A$ as the relative difference between the peak column
densities in front ($\tilde{\rho}^f$) and behind ($\tilde{\rho}^b$) the laser beam $A =
2(\tilde{\rho}^{f}-\tilde{\rho}^{b})/(\tilde{\rho}^{f}+ \tilde{\rho}^{b})$. The asymmetry $A$ is
proportional to the drag force $F$.

\begin{figure}[htbf]
    \epsfxsize=70mm \centerline{\epsfbox{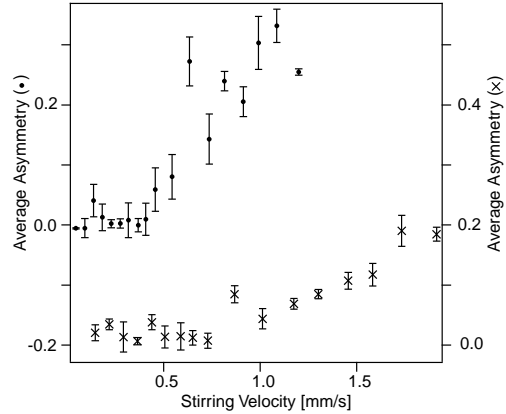}} \vspace{0.1cm}
    \caption{Density dependence
    of the critical velocity. The onset of the drag force is shown for two different condensate
    densities, corresponding to maximum sound velocities of 4.8 mm/s ($\bullet$, left axis) and 7.0
    mm/s ({\bf $\times$}, right axis). The stirring amplitudes are 29 $\mu$m and 58 $\mu$m,
    respectively.  The two vertical axes are offset for clarity. The bars represent statistical
    errors. Figure is taken from Ref.\ \protect \cite{onof00sup}. }
    \label{fig:asym_2}
\end{figure}

In Fig.\ \ref{fig:asym_2} we show measurements of the asymmetry for two maximum densities $\rho_0$
of $9 \times 10^{13}$ and $1.9 \times 10^{14}$ cm${}^{-3}$.  In each data set there is a threshold
velocity $v_c$ below which the drag force is negligible, and this threshold increases at higher
density.  Its value is close to $0.1\ c_s$ for both data sets, where $c_s$ is the sound velocity.
Above this critical velocity, the drag force increases monotonically, with a larger slope at low
density.

One can compare measurements of the asymmetry (proportional to the drag force ${\bf F}$) with the
rate of energy transferred to the condensate, ${\bf F} \cdot {\bf v}$, using the calorimetric
technique introduced in \cite{rama99}. For this, the condensate was stirred for times between
100~ms and 8~s, in order to produce approximately the same final temperature. After the stirring
beam was shut off, the cloud was allowed to equilibrate for 100 ms. Using ballistic expansion and
absorption imaging, we determined the thermal fraction and thus the temperature and total energy.

\begin{figure}[htbf]
    \epsfxsize=75mm \centerline{\epsfbox{calorimetry.eps}} \vspace{0.1cm} \caption{Calorimetry of a
    condensate. The energy transfer rate during stirring ($\bullet$, left axis) was obtained from
    temperature measurements. The error bars reflect shot-to-shot variations in the temperature. The
    results are compared to the energy transfer rate $v A(v)$ obtained from asymmetry measurements of
    the flow field during the stirring ($\times$, right axis). Figure is taken from Ref.\ \protect
    \cite{onof00sup}.}
    \label{fig:calorimetry}
\end{figure}

The calorimetric measurements can be compared with the drag force inferred from the asymmetric
density distribution. Using Eq.\ \ref{eq:drag_force}, the energy transfer rate per atom is written
in terms of the asymmetry as
\begin{equation}
    \left. {{\rm d}E \over {\rm d}t} \right|_{\rm asym}\equiv \frac{ {\bf F} \cdot {\bf v}}{N} \approx
    \frac{8}{15}\frac{\mu_0 \rho_0 l_z D}{N} v A(v)
    \label{eq:heatingasymmetry}
\end{equation}
where $D$ is the diameter of the laser beam and $l_z$ the Thomas
Fermi diameter in the radial direction.

Fig.~\ref{fig:calorimetry} shows that the calorimetric  and the
drag force measurements are in remarkable agreement over the
entire velocity range up to a single scale factor for $v A(v)$,
demonstrating the consistency between the two methods. For the
parameters of our experiment ($D \simeq 10 \mu$m, $\rho_0=1.3
\times 10^{14}$ cm${}^{-3}$, $l_z =66 \mu$m, $N= 1.8 \cdot
10^{7}$) the overall heating rate predicted by
Eq.~\ref{eq:heatingasymmetry} is 2.4 times larger than that
obtained directly from calorimetry.  Possible explanations for
this difference are the inhomogeneous density profile or a
breakdown of the quasi-static approximation embodied in Eqs.\
\ref{eq:drag_force} and \ref{eq:heatingasymmetry}.

The observed critical velocity may be related to the formation of vortices. An estimate based on
Eq.\ \ref{eq:feyn} for typical experimental parameters in sodium ($D = 10\ \mu$m, peak density
$\rho_0 = 1.5 \times 10^{14}$ cm$^{-3}$, $a = 2.75$ nm) yields $v_c \simeq 1.0$ mm/s, close to the
experimental observations. However, Eq.\ \ref{eq:feyn} depends only weakly on the speed of sound,
through the logarithmic dependence on the healing length $\xi$. In contrast, our measurements show
an approximate proportionality to the sound velocity\cite{onof00sup}, suggesting that vortex
nucleation determines the onset of dissipation.

Time-dependent simulations of the Gross-Pitaevskii equation show the formation of vortex line pairs
above a critical velocity which is close to the observed value \cite{jack00}.  Several authors have
emphasized the role of locally supersonic flow around the laser beam in the nucleation of vortices
\cite{jack00,fris92,jack99}.  In one theoretical model \cite{fris92}, the vortices are emitted
periodically at a rate that increases with velocity, and reduce the pressure gradient across the
object.  The predicted heating rate \cite{fris92,wini99} is in rough agreement with the
experimental results. Moreover, this model also predicts that the slope of the asymmetry vs.\
velocity should increase at lower density, in accord with our observations (Fig.~\ref{fig:asym_2}).

The calorimetric measurements were extended to purely thermal clouds \cite{rama00jltp}.  By
accounting for the different geometries of the clouds, we could infer the effective energy
transfer to an atom in a collision with the stirrer.  For velocities well above the critical
velocity we found almost equal energy transfer.  This result fits well into the picture that the
special properties of a condensate only show up in processes where little energy is transferred to
the atoms.

\section{Four-wave mixing of light and atoms}
\label{sec:four_wave_mixing}

Our discussion on amplification of light and atoms in a BEC in
the next section will reveal novel aspects of the coupling
between light and atoms.  In preparation for this, we want to
present some general aspects of the  scattering Hamiltonian in
Eq.\ \ref{eq:hamil} applied to the scattering of photons from the
condensate (the operators $\hat{c}^\dagger_{l}$, $\hat{c}_k $ are
now creation and annihilation operators for photons). The coupling
can be regarded as four-wave mixing of two atomic fields and two
electromagnetic fields.

When an atom is illuminated by two strong laser beams in a
$\Lambda$ configuration similar to Fig.\ \ref{fig:bogoliubov}, the
electronically excited state can be adiabatically eliminated for
sufficiently large detuning $\Delta$.   The coupling matrix
element between the two atomic ground states is $\hbar
\Omega_R/2$, where $\Omega_R$ is the two-photon Rabi frequency.
This Rabi frequency can be expressed by the (complex) electric
field strength $E_{1,2}$ of the two laser beams:
$\Omega_R=\Omega_1 \Omega_2 \cos \phi /2 \Delta$, where
$\Omega_{1,2}=E_{1,2} d/\hbar$ are the Rabi frequencies of the
individual laser beams with the atomic dipole matrix element $d$.
$\phi$ is the angle between the axes of polarization of the two
laser beams. The two-photon Rabi frequency $\Omega_R$ can be
rewritten as
\begin{equation}
  \Omega_R= d^2 E_1 E_2 \cos \phi/ 2 \hbar^2 \Delta.
  \label{eq:Rabi_freq}
\end{equation}
Using the Hamiltonian in Eq.\ \ref{eq:hamil}, the coupling matrix element squared is $|C|^2 n_1
n_2$ where $n_{1,2}$ are the photon numbers in the two beams. The two-photon Rabi frequency
$\Omega_R$ is given by
\begin{equation}
    (\hbar \Omega_R/2)^2= |C|^2 n_1 n_2.
    \label{Rabi_freq_numbers}
\end{equation}
  Expressing the number $n_{1}$ of photons with angular frequency
$\omega_0$ in a volume $V$ by the complex electric field strength $E_1$
\begin{equation}
    n_1=\epsilon_0 |E_1|^2 V/2 \hbar \omega_0,
    \label{eq:photon_number}
\end{equation}
and comparing to Eq.\ \ref{eq:Rabi_freq}  we obtain for the coupling constant $C$ between the two
modes
\begin{equation}
    C=\frac{\omega_0 d^2 \cos \phi }{2 \epsilon_0 V \Delta}.
    \label{eq:constant_C_phi}
\end{equation}

Another simple limit of the Hamiltonian in Eq.~\ref{eq:hamil} is the situation when atoms are
illuminated by a single laser beam in mode 1. Then the diagonal term $C \hat{c}^\dagger_{1}
\hat{a}^\dagger_0 \hat{c}_1 \hat{a}_0$ gives rise to the AC Stark shift. With the photon number
$n_1$, the AC Stark shift $\Delta E$ of an atom is $\Delta E =C n_1$. Using Eqs.\
\ref{eq:photon_number} and \ref{eq:constant_C_phi} with $\cos \phi =1$ one obtains the well known
result, $\Delta E =d^2 E_1^2/4 \hbar \Delta$.

Since the four-wave mixing Hamiltonian (Eq.\ \ref{eq:hamil}) applies to interactions of a
condensate both with light and atoms, we can draw analogies.  The AC Stark shift corresponds to
the mean field interaction between impurity atoms and the condensate.  In Sec.\
\ref{sec:BEC-collisions} we saw how the scattering of atoms into empty modes gave rise to the
usual elastic collision rate.  Similarly, the scattering of photons into empty modes by the
four-wave mixing Hamiltonian results in Rayleigh scattering which we want to discuss now in more
detail.

Rayleigh scattering is described by Eq.\ \ref{eq:net-rate}.  In the limit of weak scattering and
for non-interacting atoms ($S(q)=1$) one obtains
\begin{equation}
    W_+/N_0 =\frac{2 \pi}{\hbar} |C|^2 n_1 \delta(E_k- E_{k-q} - \hbar
    \omega_q^0) .
    \label{eq:Rayleigh_scattering}
\end{equation}
The scattering rate per solid angle ${\rm d} \gamma_{\rm scatt}/{\rm d}\Omega$ is given by
integrating this equation over all final states using the density of states per energy interval and
solid angle
\begin{equation}
    \frac{{\rm d}\rho}{{\rm d}E {\rm d}{\Omega}}=\frac{V \omega_0^2}{(2\pi c)^3 \hbar}.
    \label{density_of_states}
\end{equation}
Using Eqs.~\ref{eq:photon_number}, \ref{eq:constant_C_phi} one obtains
\begin{equation}
    \frac{{\rm d}\gamma_{\rm scatt}} {{\rm d}\Omega}=
    \frac{2 \pi}{\hbar} |C|^2 n_1 \frac{{\rm d}\rho}{{\rm d}E {\rm d}{\Omega}}=
    \frac{\omega_0^3 E_1^2 d^4 \cos^2 \phi}{32 \pi^2 \hbar^3 c^3 \epsilon_0
    \Delta^2} .
    \label{eq:scatt_rate_solid_angle}
\end{equation}
With the expression for the natural linewidth $\Gamma$
\begin{equation}
    \Gamma=\frac{d^2 \omega_0^3}{3 \pi \epsilon_0 \hbar c^3},
    \label{eq:gamma}
\end{equation}
this simplifies to
\begin{equation}
    \frac{{\rm d}\gamma_{\rm scatt}} {{\rm d}\Omega}=
    \frac{3 \cos^2 \phi}{8 \pi}  \frac{\Omega_1^2}{4 \Delta^2}\Gamma
    = \frac{3 \sin^2 \theta}{8 \pi}R ,
    \label{eq:scatt_rate-Rayleigh_rate}
\end{equation}
where we have defined the Rayleigh rate $R$
\begin{equation}
    R =\frac{\Omega_1^2}{4 \Delta^2}\Gamma.
    \label{eq:Rayleigh_rate}
\end{equation}
$\theta$ denotes the angle between the linear polarization of the incident light and the direction
of the scattered light.  For each scattering angle $\theta$ there are two polarizations of the
scattered light. One is orthogonal to the incident polarization, thus $\cos \phi=0$ in Eq.\
\ref{eq:constant_C_phi}, and it doesn't contribute.  The density of states in Eq.\
\ref{density_of_states} was therefore defined for only one polarization. The other polarization is
in the plane of the incident polarization and the scattering direction, thus $\cos \phi=\sin
\theta$.  $R$ was defined in Eq.\ \ref{eq:Rayleigh_rate} in such a way that integration of Eq.\
\ref{eq:scatt_rate-Rayleigh_rate} over the whole solid angle gives $\gamma_{\rm scatt}=R$.

Below, we will need another useful expression for the two-photon Rabi frequency $\Omega_R$. We use
Eq.\ \ref{Rabi_freq_numbers}, assume that one laser beam is a weak probe beam with photon number
$n_p$, express the photon number of the other (strong) beam by its Raleigh scattering rate and
obtain
\begin{equation}
    \Omega_R^{2}= 6\pi R \lambdabar^{2} c n_{\rm p} / V.
    \label{eq:Rabi_and_Rayleigh}
\end{equation}

Four-wave mixing between atoms and light has two important limiting cases. In one case, atoms are
diffracted by a standing wave of light, i.e.,\ the atoms move in the AC Stark shift potential of
two interfering light fields.  In the other case, light is diffracted by a matter wave grating, a
density modulation formed by two interfering matter waves.

In the first case, the AC Stark shift potential is proportional to $\sqrt{n_k n_l}$, where
$n_{k,l}$ are the number of photons in the two laser beams.  The diffraction efficiency is
proportional to the square of the potential, and therefore, for $N_q$ atoms in the incident mode,
the scattering rate is proportional to $n_k n_l N_q$.  In the latter case, the density modulation
caused by the interference between $N_k$ and $N_l$ atoms is proportional to $\sqrt{N_k N_l}$.  The
scattering rate for light with $n_q$ photons is then proportional to $N_k N_l n_q$.  Applying this
to a condensate with $N_0$ atoms illuminated with a (strong) laser beam with $n_k$ photons, we see
from Eq.\ \ref{eq:net-rate2} that for $N_q < n_{k-q}$  one has Bragg scattering, the scattering of
atoms from a standing wave of light.  This is usually realized by illuminating the atoms with two
laser beams.  For $N_q > n_{k-q}$ the physical picture is the diffraction of light by an atomic
density modulation.  It is this regime, which we have exploited for the amplification of atoms in a
Bose-Einstein condensate (see Sect.\ \ref{sec:Amplification of light in a dressed condensate} for a
further discussion).

Generally, the optical (or atomic) gratings are moving.  As a result, the diffracted atoms (or
photons) have an energy different from that of the incident particles.  However, one can always
transform to a moving frame where the grating is stationary and there is no energy transfer in the
scattering process.  The frequency shift due to the moving grating can therefore be regarded as
the Doppler shift related to the Gallilean transformation between the two frames.

\section{Superradiance and matter wave amplification}
\label{sec:SR_and_MWA}

Spontaneous light scattering can be stimulated when the atomic recoil state is already populated
(the $N_q$ term in Eq.\ \ref{eq:net-rate}). We have explored this process in our studies of
superradiance \cite{inou99super}, phase-coherent atom amplification \cite{inou99mwa}, and optical
amplification \cite{inou00slow}.

In all these experiments, the condensate was illuminated with a laser beam (mode $k$, also called
the ``dressing beam'').  A condensate atom scatters a photon from the laser beam into another mode
and receives the corresponding recoil momentum and energy.  Injection of atoms turns this {\it
spontaneous} process into a {\it stimulated} process and realizes an amplifier for atoms. The
injected atoms interfere with the condensate at rest and form a matter wave grating which
diffracts the dressing light. The diffraction transfers recoil momentum and energy to the atoms,
which results in a growth of the grating and therefore the number of atoms in the recoil
mode---this is the intuitive picture for atom gain. If no atoms are injected, the whole process may
start from spontaneous scattering as superradiance. We will discuss in Sec.\
\ref{sec:Amplification of light in a dressed condensate} that the build-up of the matter wave
grating can be induced also by a probe light beam resulting in optical amplification.

Eq.\ \ref{eq:net-rate} describes the scattering rate out of a condensate into a recoil state with
population $N_q$. In the limit of an empty mode for the scattered light ($n_{k-q}=0$), it reduces
to
\begin{equation}
    \label{eq:gain-rate}
    W_+ =\frac{2 \pi}{\hbar} |C|^2 N_0 n_k (N_q +1) \delta(E_k-E_{k-q} - \hbar \omega_q^B).
\end{equation}
For the high  momentum transfers considered here (on the order of
the photon recoil momentum), $S(q)=1$. Each scattering event
which transfers momentum $\hbar {\bf q}$ to the condensate,
generates a recoiling atom in mode $q$.  The final states of the
photon which are associated with a momentum transfer $\hbar
\bf{q}$ form a continuum. Integrating $W_+$ over all such final
states gives the growth rate $\dot{N_{q}}$ for the recoiling
atoms. Using Eq.\ \ref{eq:scatt_rate-Rayleigh_rate} we obtain
\begin{equation}
    \dot{N_{q}}={G_{q} (N_{q}+1) -\Gamma_{2,q} N_{q}}
    \label{eq:gain-equation}
\end{equation}
with the gain coefficient
\begin{equation}
    G_{q}=R\,N_{0}\,\frac{\sin^{2}{\theta_{q}}}{8 \pi/3}\Omega_{q}.
    \label{eq:gain-coeff}
\end{equation}

The solid angle $\Omega_{q}$ reflects the number of photon modes
which are excited together with a quasiparticle with momentum
$\hbar \bf{q}$. $N_0$ is the number of atoms in the condensate at
rest and $\theta_q$ is the angle between the polarization of the
dressing beam and the direction of photon emission. In addition,
a loss term $\Gamma_{2,q}$ was included which describes the
decoherence rate of the matter-wave grating and determines the
threshold for exponential growth. It represents the linewidth of
the two-photon process which generates recoil atoms in mode $q$.
The total scattering rate $W_{\rm tot}$ is
\begin{equation}
    W_{\rm tot}=\sum_{q} {\dot{N_{q}}}
    \label{eq:total_scattering}
\end{equation}

It is important to realize how the solid angle is divided into a sum over modes $q$.  When $N_{q}
\ll 1$, the sum in Eq.\ \ref{eq:total_scattering}  is simply an integral over the whole solid
angle and the result (without the loss term) is the Rayleigh scattering rate: $W_{\rm tot}=\Sigma
\dot{N}_j = R N_{0}$.  However, when the build-up of population $N_q$ becomes important, the
division of the solid angle into ``coherent pieces'' becomes essential. The light, emitted into the
direction $\bf k-q$, and the quasi-particles are created in the finite volume of the condensate.
Therefore, their momentum is only defined to within $\hbar$ over the dimension of the condensate.
Each ``mode'' represents a solid angle corresponding to that uncertainty (the longitudinal
momentum of the photon is determined by energy conservation, i.e.,\ the $\delta$ function in Eq.\
\ref{eq:gain-rate}). Therefore, each scattering mode spans a diffraction limited solid angle which
is approximately $\Omega_{q} \approx \lambda^2/A_q$ where $\lambda$ is the wavelength of the
scattered light and $A_q$ is the cross-section of the condensate perpendicular to the axis of
light emission.

Our Les Houches notes discuss a semiclassical derivation of the gain mechanism
\cite{stamp00leshouches}.  In this picture, one has $N_q$ atoms in the recoil mode interfering
with the condensate at rest  resulting in a modulated density.  When this atomic distribution is
illuminated with the dressing light, all the atoms can be regarded as driven oscillators. Their
radiation interferes constructively in the direction $\theta_q$ in which the density grating
diffracts the dressing light.  The phase-matching condition is fulfilled for this direction and a
solid angle $\Omega_q$ around it.  This angle can be rigorously obtained from the usual
phase-matching integral for superradiance in extended samples \cite{rehl71}
\begin{equation}
    \Omega_q = \int {\rm d} \Omega ({{\bf k}})
    \left|
    \int \tilde{\rho}({\bf r})
    \exp({\rm i} ({\bf k}_{i}- {\bf k}) \cdot {\bf r}) {\rm d}{\bf r}
    \right|^{2} ,
\end{equation}
where ${\bf k}_i$ is the wave vector of the incident light, $|{\bf k}|=|{\bf k}_{i}|$, and
$\tilde{\rho}({\bf r})$ is the normalized atomic density distribution ($\int \tilde{\rho}({\bf r})
{\rm d}{{\bf r}} =1$).

The key results of the discussion above are the gain equations, Eqs.\ \ref{eq:gain-equation} and
\ref{eq:gain-coeff}. For a condensate of cross-section $A_q$ and length $l_q$, the gain is
proportional to $N_0 \Omega_q \approx \rho_0 \lambda^2 l_q$ which is proportional to the resonant
optical density of the condensate along the direction of the scattered light.  Therefore, for an
anisotropic condensate, the gain is largest when the light is emitted along its longest axis (the
``end-fire mode'' \cite{dick64}). When the intensity of the dressing light is above threshold
($G_{q} >\Gamma_{2,q}$ in Eq.\ \ref{eq:gain-equation}), a condensate will emit highly directional
beams of light and atoms along the direction for which the gain is highest.  Our observation of
this phenomenon has been described in Refs.\ \cite{stamp00leshouches,inou99super}.

\begin{figure}
    \begin{center}
    \includegraphics[height=2.5in]{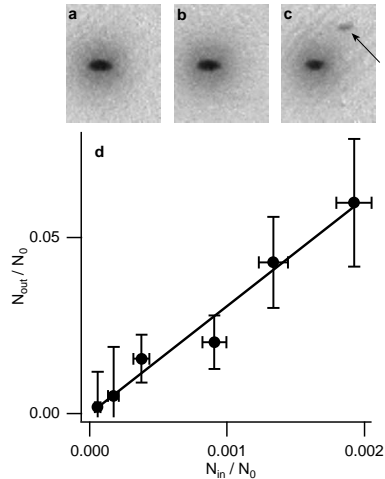}
    \caption{Input--output characteristic of the matter-wave amplifier.  ({\bf a-c}) Typical
    time-of-flight absorption images demonstrating matter wave amplification. The output of the seeded
    amplifier ({\bf c}) is clearly visible, whereas no recoiling atoms are discernible in the case
    without amplification ({\bf a}) or amplification without the input ({\bf b}). The size of the
    images is 2.8~mm $\times$ 2.3~mm. ({\bf d}) Output of the amplifier as a function of the number of
    atoms at the input.  A straight line fit shows a number gain of 30.  Reprinted by permission from
    Nature, Ref.~\protect \cite{inou99mwa}, copyright 1999 Macmillan Magazines Ltd.
    \label{fig:MWAgain}}
    \end{center}
\end{figure}

The amplification of atoms is conceptionally even simpler:  Atoms in a certain recoil mode $q$ are
injected into the condensate and are amplified with a gain coefficient $G_q$.  We only summarize
the main experimental results---for a full account see Refs.\ \cite{stamp00leshouches,inou99mwa}.

\begin{figure}[btp]
    \begin{center}
    \includegraphics[height=2in]{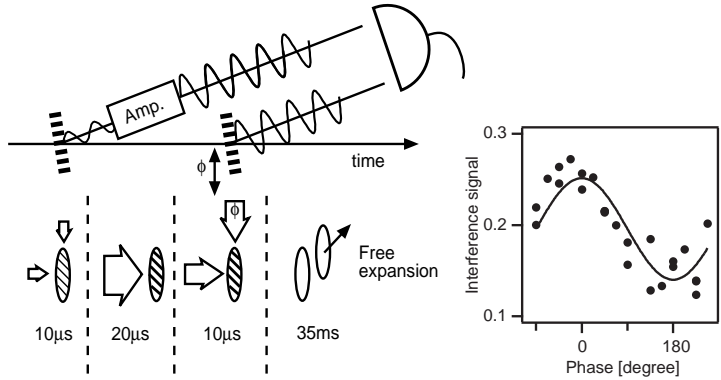}
    \caption{ Experimental scheme for observing phase coherent matter wave amplification. A
    small-amplitude matter wave was split off the condensate by applying a pulse of two off-resonant
    laser beams (Bragg pulse). This input matter wave was amplified by passing it through the
    condensate pumped by a laser beam. The coherence of the amplified wave was verified by observing
    its interference with a reference matter wave, which was produced by applying a second (reference)
    Bragg pulse to the condensate.  The total number of atoms in the recoil mode showed
    constructive and destructive interference between the amplified input and the reference matter wave
    as the phase of the reference wave was scanned.
    Reprinted by permission from Nature, Ref.~\protect \cite{inou99mwa},
    copyright 1999 Macmillan Magazines Ltd.
    \label{fig:MWAsetup}}
    \end{center}
\end{figure}

Input matter waves with momentum $\hbar {\bf q}$ were produced by exposing the condensate to two
laser beams.  Their difference frequency was tuned to the resonance for Bragg scattering (see
Sec.~\ref{sec:Bragg_spect}). The stimulated  redistribution of photons among the two beams
transferred recoil momentum to the atoms. The fraction of atoms in the recoil state was controlled
by the intensity and duration of the Bragg pulse. While these input atoms were still in the
condensate volume, they were amplified when the condensate was exposed to the dressing beam.
Fig.~\ref{fig:MWAgain} shows the input-output characteristics of the amplifier. The gain was
controlled by the intensity of the pump pulse (see Eq.~\ref{eq:gain-coeff}) and typically varied
between 10 and 100. Fig.~\ref{fig:MWAgain}d shows the observed linear relationship between the atom
numbers in the input and the amplified output with a number gain of 30.

This atom amplifier is a narrow band amplifier.  It only amplifies input momentum states which can
be populated by condensate atoms by scattering a photon of the dressing beam.  The possible input
states lie on a sphere in momentum space.  Its center is displaced from the origin (the momentum of
the condensate at rest) by the momentum $\hbar {\bf k}$ of the dressing beam.  The thickness of the
momentum sphere is determined by the momentum  uncertainty of the condensate, which is $\hbar$ over
its size, as was directly measured using Bragg spectroscopy \cite{sten99brag}.  In our experiment,
the input momentum was automatically matched to the amplifier's narrow bandwidth since the input
beam was created by Bragg scattering, and one of the Bragg beams was identical to the dressing
beam.

The Hamiltonian in Eq.\ \ref{eq:hamil} provides {\it coherent} coupling between the light and
atoms.  Therefore, the atom amplification should be phase-coherent. This was experimentally
verified with an interferometric technique. For this, a reference matter wave was split off the
condensate in the same way as the first (input) wave (see Fig.\ \ref{fig:MWAsetup}). The phase of
the reference matter wave was scanned by shifting the phase of the radio-frequency signal that
drove the acousto-optic modulator generating the axial Bragg beam. We then observed the
interference between the reference and the amplified matter waves by measuring the number of atoms
in the recoil mode.

\section{Amplification of light in a dressed condensate}
\label{sec:Amplification of light in a dressed condensate}

A dressed condensate (Fig.\ \ref{fig:setup-light-amp}), a condensate illuminated by laser light,
was used as an atom amplifier.  Now we develop this picture further. Rayleigh scattering produces
scattered photons and recoiling atoms. In the dressed atom picture, this is described as the decay
of the dressed condensate  into a photon and recoiling atom, or in other words, the dressed
condensate can spontaneously emit pairs of photons and atoms. The amplification of atoms discussed
in the previous sections solely focuses on the recoiling atoms ``emitted'' by the dressed
condensate. Although recoiling atoms and scattered photons are emitted in pairs, the photons leave
the condensate almost instantaneously and there is no significant population build-up. Formally, as
discussed in the Les Houches notes, one can adiabatically eliminate the light field from coupled
equations and obtain the gain equation for the matter waves.

On the other hand, the dressed condensate should act also as an amplifier for light.  An input
optical field should stimulate Rayleigh scattering processes which results in  photons scattered
into the input mode. Our recent experiments on optical amplification in a BEC \cite{inou00slow}
required a more general description of the interplay between optical and matter wave amplification.

On the following pages, we present a general discussion of four-wave mixing of light and atoms.  We
first start simply with the gain cross-section for the light and a complex index of refraction. The
recoiling atoms enter the picture in two stages, first within the framework of Heisenberg equations
in the undepleted-pump approximation, and then using optical Bloch equations.

\subsection{Cross-section for optical gain and slow light}

\begin{figure}[htbf]
    \epsfxsize=69mm \centerline{\epsfbox{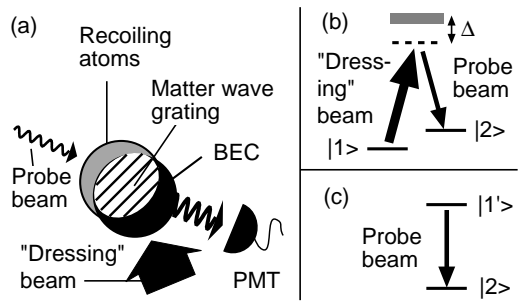}}\vspace{0.5cm}
    \caption{Amplification of light and atoms by off-resonant light scattering. (a) The fundamental
    process is the absorption of a photon from the ``dressing'' beam by an atom in the condensate
    (state $|1 \rangle$), which is transferred to a recoil state (state $|2 \rangle$) by emitting a
    photon into the probe field. The intensity in the probe light field was monitored by a
    photomultiplier. (b) The two-photon Raman-type transition between two motional states
    ($|1\rangle, |2\rangle$) gives rise to a narrow resonance. (c) The dressed condensate is the
    upper state ($|1^{\prime}\rangle$) of a two-level system, and decays to the lower state
    (recoil state of atoms, $|2\rangle$) by emitting a photon. Figure is taken from Ref.\
    \protect\cite{inou00slow}}
    \label{fig:setup-light-amp}
\end{figure}

The physical picture behind the optical gain of the dressed
condensate is as follows: if a very weak probe beam is injected
into the dressed condensate, it acts together with the dressing
beam as a pair of Bragg beams and creates recoiling atoms. This
process transfers photons from the dressing beam into the probe
beam. At higher gain, the recoiling atoms become significant. They
move out of the condensate (or decohere) on a time scale
$\Gamma_2^{-1}$ which is the inverse of the linewidth of the
Bragg transition. In steady state, the number of recoiling atoms
$N_q$ in the volume of the condensate is proportional to the
intensity of the probe light. Those recoiling atoms interfere
with the condensate at rest and form a diffraction grating which
diffracts the dressing beam into the path of the probe light
resulting in amplification of the probe light (Fig.\
\ref{fig:setup-light-amp}).

An expression for the gain can be derived in analogy to a fully inverted two-level system with
dipole coupling which would have a gain cross-section of $6\pi\lambdabar^{2}$ for radiation with
wavelength $\lambda (= 2\pi\lambdabar)$. For the Raman-type system in Fig.\
\ref{fig:setup-light-amp}b, the gain is reduced by the excited state fraction, $R / \Gamma$ (where
$R$ is the Rayleigh scattering rate for the dressing beam and $\Gamma$ is the linewidth of the
single-photon atomic resonance) and increased by $\Gamma/\Gamma_{2}$, the ratio of the linewidths
of the single-photon and two-photon Bragg resonances. Thus the expected cross-section for gain is
\begin{equation}
    \sigma_{\rm gain}=6\pi\lambdabar^{2} \frac{R}{\Gamma_{2}}.
    \label{eq:twophotongain}
\end{equation}

The lineshape of the optical gain is that of the two-photon Bragg resonance. Due to the long
coherence time of a condensate, it has a very narrow linewidth.  Such a narrow band gain is
accompanied by a slow group velocity of light. This can be described by a complex index of
refraction $n(\omega)=n_1(\omega) + {\rm i} n_2(\omega)$.

For a Lorentzian resonance curve with FWHM of $\Gamma_2$ for the
gain, the complex index of refraction is
\begin{equation}
    n(\omega)=n_{\infty}+\frac{g^\prime}{\delta + {\rm i}}=
    n_{\infty}+\frac{g^\prime \delta}{1+\delta^2} - \frac{g^\prime {\rm i}}{1+\delta^2}
    \label{eq:complex_index}
\end{equation}
where $n_{\infty}$ is the background index of refraction,
$g^\prime$ denotes the strength of the resonance, and
$\delta=(\omega-\omega_0)/(\Gamma_2 /2)$ is the normalized
detuning from the resonance at $\omega_0$. The imaginary part of
Eq.\ \ref{eq:complex_index} has the usual Lorentzian lineshape.

The real and imaginary parts of the index of refraction are connected by Kramers-Kronig relations.
For the special case above of a Lorentzian lineshape the gain and dispersion at resonance are
connected by
\begin{equation}
    \frac{d n_1}{d\omega}\Bigr|_{\omega=\omega_0} = - \frac{2}{\Gamma_2} n_2
    \Bigr|_{\omega=\omega_0}.
    \label{eq:real_complex_part}
\end{equation}
A steep slope of the (real part of) the index of refraction gives
rise to a slow group velocity of light
\begin{equation}
    {v_{g}}=\frac{c}{\omega ({\rm d} n_1/{\rm d}\omega)+ n_1}.
    \label{eq:vg}
\end{equation}

Eqs.\ \ref{eq:real_complex_part} and \ref{eq:vg} imply a simple relationship between the gain and
delay time for an optical pulse.  The amplitude of an optical pulse at frequency $\omega_0$ which
propagates through a medium with the index of refraction $n$ of length $l$ is amplified by a factor
\begin{equation}
    g=\exp(-n_2 \omega_0 l/c).
    \label{eq:g_with_exp}
\end{equation}
When the first term in the denominator of Eq.\ \ref{eq:vg} is dominant, the delay time $\tau_{D}$
of the pulse is
\begin{equation}
    \tau_{D} = \frac{l}{v_{g}} \approx l \frac {\omega_0}{c} \frac{{\rm d} n_1}{{\rm d}\omega}
    =\frac{2 \ln g }{\Gamma_{2}}.
    \label{eq:delay_gain}
\end{equation}
This equation provides a simple relationship between a narrow
band gain and pulse delay~\cite{casp71yariv}.  A non-inverted
absorptive two-level system gives rise to ``superluminal'' pulse
propagation~\cite{chu82}.

For the experimental study of the optical gain, a condensate was
illuminated (``dressed'') with a single off-resonant laser beam
and probed with another laser beam, which was red-detuned by 91
kHz to satisfy the Bragg resonance condition. Both the dressing
beam and the probe beam were in the plane perpendicular to the
long axis of the condensate, and intersected at an angle of 135
degrees. The probe beam, which propagated parallel to the axis of
imaging, was much larger than the condensate size. In order to
block all the light that did not pass through the condensate, a
slit was placed at an intermediate imaging plane. The light
transmitted by the slit was recorded with a photomultiplier. The
polarization of each beam was set parallel to the long axis of
the condensate to suppress superradiance to other recoil modes
\cite{inou99super}.

\begin{figure}[htbf]
    \epsfxsize=69mm \centerline{\epsfbox{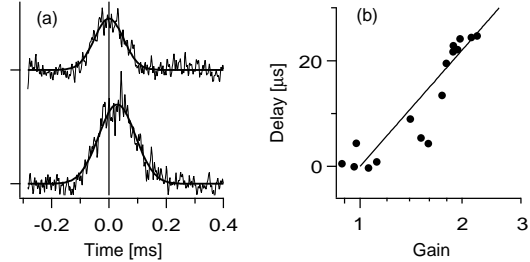}} \vspace{0.5cm}
    \caption{Pulse delay due to
    light amplification. (a) About 20 $\mu$s delay was observed when a Gaussian pulse of about 140
    $\mu$s width and 0.11 mW/cm$^{2}$ peak intensity was sent through the dressed condensate (bottom
    trace). The top trace is a reference taken without the dressed condensate. Solid curves are
    Gaussian fits to guide the eyes. (b) The observed delay was proportional to the logarithm of the
    observed gain. Figure is taken from Ref.\ \protect\cite{inou00slow}.}
    \label{fig:gain-delay}
\end{figure}

Fig.\ \ref{fig:gain-delay} shows that light pulses were delayed by about 20 $\mu$s across the 20
$\mu$m wide condensate corresponding to a group velocity of 1 m/s.  This is one order of magnitude
slower than any value reported previously (see Ref.\ \cite{hau99} and references therein). Fig.\
\ref{fig:gain-delay}b presents the experimental verification of the relationship between gain and
delay time (Eq.\ \ref{eq:delay_gain}).

\subsection{Relation between optical gain and atomic gain}

Both the optical gain $g$ (Eqs.~\ref{eq:twophotongain} and \ref{eq:g_with_exp}) and the matter wave
$G$ gain (Eq.\ \ref{eq:gain-coeff}) have the same origin, stimulated Rayleigh scattering.
Therefore, the two gain coefficients should be related. The expression for $G$ (Eq.\
\ref{eq:gain-coeff}) involves a solid angle factor $\Omega_q$ which is proportional to
$\lambda^2/A$.  To be consistent with the optical Bloch equation (to be discussed below) we use now
$2 \lambda^2/A$ for $\Omega_q$ and obtain
\begin{equation}
    G=\frac{R N_0}{A} \frac{3}{4\pi}\lambda^2=(\rho_0 l \sigma_{\rm gain}/2) \Gamma_2.
    \label{eq:gain1}
\end{equation}
The gain $g$ for the amplitude of the optical field is
\begin{equation}
    g=\exp(\rho_0 l \sigma_{\rm gain}/2)=  \exp(G/\Gamma_2) \approx 1+\frac{G}{\Gamma_2}
    \label{eq:optical_gain}
\end{equation}
with the last equation being an approximation for small gain.  However, Eq.\ \ref{eq:optical_gain}
cannot be universally valid. When the gain $G$ is above the threshold for superradiance,
$G>\Gamma_2$ (Eq.\ \ref{eq:gain-equation}) the optical gain should diverge: a single recoiling
atom created by the probe light and dressing light is exponentially amplified and creates a huge
matter wave grating which will diffract the dressing light into the probe light path, thus
amplifying the probe light by a divergent factor. Indeed, as we will derive below, Eq.\
\ref{eq:optical_gain} is only valid at small values of $G/\Gamma_2$.

If we can neglect the depletion of the dressed condensate and the dressing laser beam we can
simplify the interaction Hamiltonian in Eq.\ \ref{eq:hamil} to
\begin{equation}
    {\cal H^\prime} = C^\prime (\hat{a}^{\dagger} \hat{c}^{\dagger}+\hat{a} \hat{c}).
    \label{eq:simple_ham}
\end{equation}
Here, $\hat{a} (\hat{c})$ indicates the atomic (light) field to be amplified, but the following
derivation is completely symmetric between the two fields.  This Hamiltonian is a standard
down-conversion Hamiltonian.  Here it describes the down-conversion of the dressed condensate into
photons and recoiling atoms.  Considering only two modes neglects propagation effects in the
amplification.  The coefficient $C^\prime$ defines the time constant of the amplification and is
proportional to the amplitude of the dressing beam and the square root of the number of atoms in
the condensate. The Heisenberg equations of motions are
\begin{eqnarray}
    {\rm i} \dot{\hat{a}} &=& [\hat{a},{\cal H^\prime}] = C^\prime \hat{c}^{\dagger},\\
    {\rm i} \dot{\hat{c}^{\dagger}} &=& [\hat{c}^{\dagger},{\cal H^\prime}]=- C^\prime \hat{a}.
\end{eqnarray}
This leads to exponential growth of $\hat{a}$ and $\hat{c}$ (proportional to $\exp(C^\prime t)$).
However, to describe the physical situation in the experiments, one has to allow for damping by
introducing $\Gamma_a (\Gamma_c)$ as phenomenological damping time constants for
$\hat{a}$($\hat{c}$).  We also include source terms (input fields) and approximate the operators
by $c$ numbers:
\begin{eqnarray}
    \dot{a} &=& -\frac{\Gamma_a}{2} (a-a_0) - {\rm i}C^\prime c^*, \nonumber \\
   \dot{c^*} &=& {\rm i} C^\prime a -\frac{\Gamma_c}{2} (c^*-c^*_0).
   \label{eq:coupled_eq_damp}
\end{eqnarray}
The solutions show relaxation ($C^{\prime 2}\le \Gamma_a \Gamma_c/4$) or exponential growth
($C^{\prime 2}\ge \Gamma_a \Gamma_c/4$) depending on the strength of the coupling relative to the
damping rates. The ``gain'' below the threshold can be defined as $a(t \rightarrow \infty)/a_0$ for
atoms (assuming $c_0=0$) and as $c(t \rightarrow \infty)/c_0$ for light (assuming $a_0=0$),
yielding
\begin{equation}
    g=\frac{a(t \rightarrow \infty)}{a_0} = \frac{ c(t\rightarrow \infty)}{c_0}
    = \frac{ {\Gamma_a \Gamma_c}/ {4}}{({\Gamma_a \Gamma_c}/4) - C^{\prime 2}}.
    \label{eq:two_gains}
\end{equation}
The fact that the two gain coefficients are equal is a general property of parametric amplification
where two kinds of particles are produced in pairs.

In the limiting case that one field is strongly damped (e.g.\ that light quickly escapes from the
system, $\Gamma_{c} \gg \Gamma_{a}$), one can adiabatically eliminate this field from the coupled
equation (assuming no photon input ($c^*_0=0$))
\begin{equation}
    c^*= \frac{2 {\rm i} C^\prime}{\Gamma_c} a
    \label{eq:ad_elim}
\end{equation}
and obtain a single gain equation for $a$. The gain equation for the atom field is
\begin{equation}
    \dot{a} = -\frac{\Gamma_a}{2} (a-a_0) + \frac{2 C^{\prime 2}}{\Gamma_c} a.
    \label{eq:single_eq}
\end{equation}
In the absence of damping, the atom number would increase exponentially with a rate constant $4
C^{\prime 2}/\Gamma_c$ which we therefore identify with the atom gain rate coefficient $G$ in
Eq.~\ref{eq:gain1}. This can be shown explicitly using $C^{\prime 2}= |C|^2 N_0 n_k/\hbar^2$ and
setting the mode volume $V=A L$, where $A$ is the cross section of the condensate.  The axial
length $L$ could be the condensate length $l$, but will cancel out.  Eqs.~\ref{Rabi_freq_numbers}
and \ref{eq:Rabi_and_Rayleigh} yield $4 C^{\prime 2}/\Gamma_c = 2G c/L \Gamma_c$ which equals $G$
when we set the decay rate $\Gamma_c/2$ equal to the photon transit time $c/L$. We can then rewrite
the gain calculated above as
\begin{equation}
    g=\frac{\Gamma_{a}}{\Gamma_{a}-{4 C^{\prime 2}}/\Gamma_{c}}=
     \frac{\Gamma_{a}}{\Gamma_{a}-G}.
\end{equation}
For the dressed condensate, we identify $\Gamma_a$ with $\Gamma_2$.  As expected, at the threshold
to superradiance ($G=\Gamma_{2}$), the (steady-state) gain for both light and matter waves
diverges. The gain can be rewritten as
\begin{equation}
    g=\Gamma_2/(\Gamma_2-G) = 1 + G/(\Gamma_2-G).
    \label{eq:optical_gain2}
\end{equation}
In the low gain limit, this yields the same result as Eq.~\ref{eq:optical_gain}.  The comparison
with Eq.~\ref{eq:optical_gain} shows that the effect of the coupled equations is to replace the
two-photon linewidth $\Gamma_2$ in Eq.\ \ref{eq:optical_gain} by the dynamic coherence decay rate
$\Gamma_2-G$. Since propagation effects have been excluded, we can't expect to obtain the
exponential factor in Eq.~\ref{eq:optical_gain}, but rather the linearized form. The expansion
\begin{equation}
    g= 1 + (G/\Gamma_2) + (G/\Gamma_2)^2+ \ldots
    \label{eq:opt_gain_expansion}
\end{equation}
describes the transition from (linear) single-atom gain to (nonlinear) collective gain.

\subsection{Optical Bloch equations}

The discussion in the previous two sections assumed that the condensate is undepleted---i.e.,\ we
calculated properties of a condensate with all the atoms in the initial dressed state. However, the
presence of the dressing light and the probe light depletes the condensate.  Furthermore, the
calculated amplification coefficients are only valid in a quasi-steady state regime which is
usually preceded by transient behavior.  A correct interpretation of the experimental results
required a more complete description of the dynamics of the system which will be developed in this
section using optical Bloch equations.

We proceed in two steps. In the limit of weak optical gain (or strong probe laser intensity), we
will use the ordinary optical Bloch equations where the laser fields are treated as constant.
Later we will introduce an additional equation for the dynamics of the probe light.  The
condensate at rest ($|1 \rangle$) and the atoms in the recoil state ($|2 \rangle$) are treated as a
two-level system coupled by the four-wave mixing Hamiltonian which gives rise to a two-photon Rabi
frequency $\Omega_R$ (Eq.\ \ref{eq:Rabi_freq}). The coherence between those two states decays at a
rate $\Gamma_{2}/2$. Assuming constant $\Omega_R$, the optical Bloch equations at resonance take
the following simple form
\begin{eqnarray}
    \dot{v}&=&-\frac{\Gamma_{2}}{2} v - \Omega_R w \label{vdot1}\\
    \dot{w}&=&\Omega_R v
    \label{wdot}
\end{eqnarray}
where $v= 2 \;{\rm Im} (\rho_{12})$ represents the amplitude of
the matter wave grating ($\rho_{ij}$ is the atomic density
matrix) and $w = \rho_{22} - \rho_{11}$ is the population
difference between the two states~\cite{cohe92}.

The eigenvalues of the matrix
\begin{equation}
    \pmatrix{
    -\Gamma_{2}/2& -\Omega_R \cr
    \Omega_R & 0 \cr
    }
\end{equation}
are $\lambda_{\pm} = -\Gamma_{2}/4 \pm \sqrt{ (\Gamma_{2}/4)^{2} - \Omega_R^{2}}$.  In the limits
of large and small laser intensities one obtains
\begin{equation}
    \lambda_{\pm}= \cases{ -\frac{\Gamma_2}{4} \pm {\rm i} \Omega_R & $\Gamma_{2}/4 \ll \Omega_R$ \cr
    -\frac{\Gamma_{2}}{2}, -\frac{2 \Omega_R^{2}}{\Gamma_{2}}& $\Gamma_{2}/4 \gg \Omega_R$ \cr }.
\end{equation}
This means that at high intensities the system exhibits damped oscillations---Rabi oscillations
between the two levels.  At low intensities, there is relaxation in two steps:  The coherence is
damped with a rate of $\Gamma_{2}/2$, followed by depletion of atoms in the condensate, which
happens at a rate of $2 \Omega_R^{2}/\Gamma_{2}$. It is in this temporal window ($2/\Gamma_{2} < t
< \Gamma_{2}/2\Omega_R^{2}$) that the perturbative treatment with the complex index of refraction
applies. For longer times, the condensate becomes depleted and the assumption that most of the
atoms are in the initial dressed state is no longer valid.

\begin{figure}[htbf]
    \epsfxsize=69mm \centerline{\epsfbox{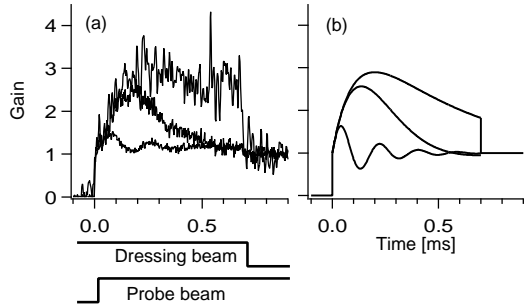}}\vspace{0.5cm}
    \caption{Gain and
    temporal behavior of light pulses propagating through a dressed condensate. (a) Observed probe
    pulse output from a dressed condensate. The probe light intensities were $5.7 \,{\rm
    mW/cm}^{2}$ (bottom), $1.5 \,{\rm mW/cm}^{2}$ (middle), $0.10 \,{\rm mW/cm}^{2}$ (top), while
    the dressing beam intensity was $5 \, {\rm mW/cm}^{2}$, which was just below the threshold for
    superradiance. The plotted signals were normalized by the incident probe intensity and show
    the gain for the probe light. (b) Calculated probe light output for typical experimental
    parameters. Rabi oscillations develop into steady state gain as the intensity of the probe
    light is reduced. Figure is taken from Ref.\ \protect\cite{inou00slow}.}
    \label{SquarePulse}
\end{figure}

The optical Bloch equations can be analytically solved for a step function input. With the initial
condition that at time $t=0$ all the atoms are in the condensate at rest ($w(t=0)=-1$, $v(t=0)=0$)
one obtains
\begin{equation}
    v(t) = \cases{ \frac{\Omega_R}{\sqrt{\Omega_R^2-({\Gamma_2}/{4})^2}} \exp (-\frac{\Gamma_2}{4} t )
    \sin \left(\sqrt{\Omega_R^2-({\Gamma_2}/4)^2} \, t \right) & $\Omega_R \ge \frac{\Gamma_2}{4}$ \cr
    \frac{\Omega_R}{\sqrt{({\Gamma_2}/4)^2-\Omega_R^2}} \exp(-\frac{\Gamma_2}{4} t )\sinh
    \left(\sqrt{({\Gamma_2}/4)^2-\Omega_R^2} \, t \right) & $\Omega_R \le \frac{\Gamma_2}{4}$ \cr }
    \label{v_general}
\end{equation}
simplifying in the limit of small probe laser intensity
($\Omega_R \ll \Gamma_2/4$) to
\begin{eqnarray}
    v(t)&\approx&  \frac{2 \Omega_R}{\Gamma_2} \left(-\exp(\frac{-\Gamma_2}{2} t)
    + \exp (- \frac{2 \Omega_R^2}{\Gamma_2} t )\right) \label{approx_v1} \\
    &=&  \frac{2 \Omega_R}{\Gamma_2} \left(-\exp(-\frac{\Gamma_2}{2} t) + 1\right) \qquad t \ll
    \Gamma_2 / \Omega_R^2.
    \label{approx_v2}
\end{eqnarray}
By reducing the probe power, the Rabi oscillations slow down and become overdamped and a
(quasi-)steady state gain is obtained. Inserting Eq.\ \ref{approx_v1} into Eq.\ \ref{wdot} one
obtains the transition rate $N_0 \dot{w}/2=N_0 \Omega_R v/2$ which is the number of photons per
unit time emitted by the dressed condensate.  To obtain the gain one has to normalize by the input
photon flux $c n_p /l$ where $n_p$ is the number of photons in the condensate.  The amplitude gain
is then (assuming small gain)
\begin{equation}
    g = 1 + \frac{N_0 \Omega_R  l}{4 c n_p} v.
    \label{gain_OBE}
\end{equation}
Using the asymptotic behavior of Eq.\ \ref{approx_v2} ($v(t)
\approx 2 \Omega_R/\Gamma_2$), Eq.\ \ref{eq:Rabi_and_Rayleigh}
for $\Omega_R$ and Eq.\ \ref{eq:twophotongain} one obtains $ g
 = 1 + \rho_0 \sigma_{\rm gain} l / 2$ which agrees with Eqs.\
\ref{eq:optical_gain} and \ref{eq:optical_gain2} in the
low-intensity limit. Eq.\ \ref{gain_OBE} thus has the correct
asymptotic limit, but it also describes transient behavior when
the general solution for $v(t)$ (Eq.\ \ref{v_general}) is used.
Theoretical traces based on Eq.\ \ref{gain_OBE} are directly
compared to the experimental results in Fig.\ \ref{SquarePulse}.

In the experiment, we used long square probe pulses for the probe light (Fig.\ \ref{SquarePulse}).
When the dressing beam was suddenly switched off, a sudden change in the observed probe light
intensity was evidence for optical gain. At the lowest probe intensity, the depletion of atoms in
the condensate was negligible and a clear step at the switch off was observed, corresponding to a
gain of $\approx 2.8$. The initial rise time of $\approx 100 \, \mu s$ is the coherence time of the
dressed condensate. At high probe laser power we observed Rabi oscillations in the transmitted
probe light. Note that all the traces were normalized by the probe beam intensity, and the
oscillatory trace at the bottom was obtained at the highest probe beam intensity. The oscillations
reflect simple two-level Rabi oscillations of atoms between the two motional states driven by the
two-photon Bragg coupling.

\begin{figure}
    \begin{center}
    \epsfig{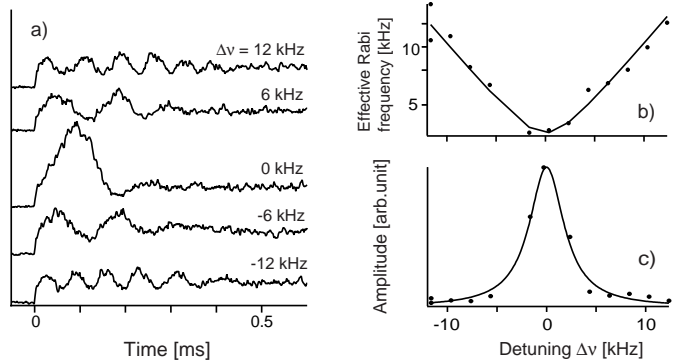} \caption{Rabi oscillations of
    atoms
    observed by monitoring the transmitted probe light intensity. a) Rabi oscillations for
    different detunings of the probe light from the two-photon resonance. b) Observed frequencies
    of Rabi oscillation versus detuning. The solid line is a fit to Eq.\ \ref{eq:effrabiosci}.  c)
    Amplitude of the oscillations versus detuning. The solid line is a Lorentzian fit with a
    linewidth of about 4 kHz.} \label{fig:detuned_rabi_osc}
    \end{center}
\end{figure}

When the probe laser frequency was detuned from the two-photon resonance, the frequency of the
Rabi oscillations increased (Fig.\ \ref{fig:detuned_rabi_osc}).  Optical Bloch equations with
detuning \cite{cohe92} predict oscillations at
\begin{equation}
    \Omega_{\rm eff}=\sqrt{\Omega_R^2 + \Delta\omega^2}
    \label{eq:effrabiosci}
\end{equation}
where $\Delta\omega$ is the detuning, in agreement with observations.

For large optical gain, the Rabi frequency $\Omega_R$ increases during the pulse and the above
treatment is no longer valid. Therefore, we derive now a second equation which treats the Rabi
frequency as a dynamic variable.  The population transfer to the recoil state ($\dot{w}$) results
in an increase of the number of the probe beam photons inside the condensate volume
\begin{equation}
\dot{n_p}= c (n_p^{0}-n_p) /l + N_{0} \dot{w}/ 2 ,
\label{eq:number_increase}
\end{equation}
where $l$ is the length of the condensate with $N_{0}$ atoms and $cn_p^{0}/l$ is the input photon
flux. Without gain, the steady-state number of photons in the condensate volume would be $n_p^0$.
Eq.~\ref{eq:number_increase} neglects propagation effects of the light by replacing the
non-uniform electric field by an average value (such a ``mean-field model'' is only qualitative,
see \cite{gros82}).

Replacing the photon number by the Rabi frequency (Eq.~\ref{eq:Rabi_and_Rayleigh}) leads to
\begin{equation}
    2 \Omega_R \dot{\Omega_R}=\frac{c}{l} (\Omega_0^2-\Omega_R^2) +
    \frac{N_0}{2}  \frac{R 6\pi\lambdabar^{2} c}{V}  \Omega_R v
\end{equation}
where $\Omega_0$ is the two-photon Rabi frequency due to the input probe beam and the dressing
beam. For small gain, we approximate $\Omega_0^2-\Omega_R^2 \approx 2 \Omega_R(\Omega_0-\Omega_R)$.
This approximation should retain the qualitative features of the coupled light-atom system even
when the small gain approximation is no longer quantitative. Indeed, we will obtain results
consistent with our previous treatment (Eq.\ \ref{eq:optical_gain2}) which was not limited to small
gain.  Using Eq.\ \ref{eq:gain1} for the atom gain $G$ we obtain
\begin{equation}
    \dot{\Omega_R}=\frac{c}{l} (\Omega_0 - \Omega_R + \frac{G}{2} v )
    \label{omegadot}
\end{equation}

This equation together with Eqs.~\ref{vdot1} and \ref{wdot} forms
a set of coupled equations describing the combined dynamics of
the atom and light fields. The situation is analogous to the
optical laser, where the atomic polarization and the electric
field inside the cavity are coupled. However, the role of atoms
and light is reversed: in the optical laser, the cavity lifetime
is usually longer than the coherence time of the atomic
polarization, whereas in our case the extremely long coherence
time of the condensate dominates. This would correspond to the
bad cavity limit of the optical laser which is usually not
realized (see \cite{vane95} and references therein).

Assuming rapid relaxation of the light field ($\dot{\Omega_R}=0$ in Eq.\ \ref{omegadot}) leads to
\begin{equation}
    \Omega_R = \Omega_0 +\frac{G}{2} v.
    \label{omega_eq}
\end{equation}
Inserting this into Eqs.\ \ref{vdot1} and \ref{wdot} adiabatically eliminates the light field.
This treatment is more general than the Heisenberg equations above, where we had neglected
condensate depletion.  To check for consistency, we now assume an undepleted condensate ($w = -1$)
and obtain
\begin{equation}
    \dot{v} = \frac{G-\Gamma_{2}}{2} v + \Omega_0.
    \label{vdot2}
\end{equation}
Below the threshold for superradiance, ($G\le \Gamma_{2}$), $v$ relaxes with a time constant of
$2/(\Gamma_{2}-G)$ to $v=2 \Omega_0/(\Gamma_2-G)$.  This and Eq.\ \ref{omega_eq} show that the gain
$g$ for the probe beam is
\begin{equation}
    g= 1+\frac{G} {\Gamma_2-G}
    \label{gain}
\end{equation}
in agreement with Eq.\ \ref{eq:optical_gain2}.

\subsection{Optical probe of matter wave gain}

The matter wave grating formed inside the condensate is responsible for both atomic and optical
gain.  We now briefly describe experiments where the dynamics of the matter wave grating could be
directly observed by monitoring the probe light. We first created a matter wave grating with a
Bragg pulse and then observed its time evolution by monitoring the diffracted dressing beam. The
initial seed pulse was 100 $\mu$s long and transferred about 5\% of the atoms to the recoil state.

\begin{figure}[htbf]
\epsfxsize=79mm \centerline{\epsfbox{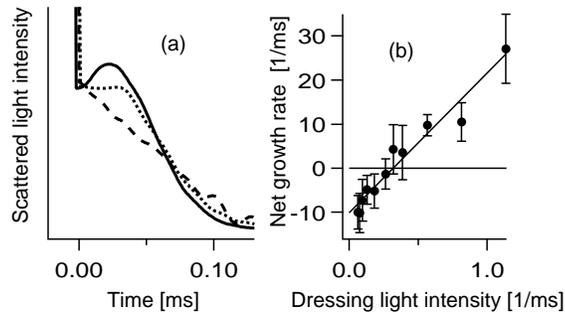}}\vspace{0.5cm} \caption{Pump-probe
spectroscopy of a matter wave grating inside the condensate. (a) First, approximately 5~\% of atoms
were transferred to the recoil state by the two-photon Bragg transition. Then the dynamics of the
matter wave grating was observed {\it in situ} by illuminating the grating with off-resonant
dressing light and monitoring the diffracted light intensity. All traces were normalized to the
same diffracted light intensity at $t=0$. The dressing beam intensity was $2.9 \, {\rm
mW/cm^{2}}$(bottom), $5.7 \, {\rm mW/cm^{2}}$(middle), $13 \, {\rm mW/cm^{2}}$(top). (b) The
initial growth rate of the grating vs.\ light intensity shows the threshold for net gain. The
intensity of the dressing beam is given in units of the single-atom Rayleigh scattering rate.
Figure is taken from Ref.\ \protect\cite{inou00slow}.} \label{pump_probe}
\end{figure}

At lower intensities for which atom amplification was negligible, the grating showed a simple decay
(Fig.\ \ref{pump_probe}). At higher intensities, collective gain started to compensate the loss,
and at intensities above a threshold, net amplification was observed.  The initial growth rate
(Fig.\ \ref{pump_probe}) followed the linear dependence on the intensity of the dressing beam
($\propto (G-\Gamma_2$)) predicted by Eq.\ \ref{vdot2} and Refs.~\cite{inou99super,moor99super}.
The net growth of the matter wave grating was studied previously by observing an increase in the
number of recoiling atoms in time-of-flight images~\cite{inou99mwa}, whereas Fig.~\ref{pump_probe}
was obtained by monitoring the dynamics of amplification {\it in situ} by observing light instead
of atoms.

Extrapolating the decay rate in Fig.\ \ref{pump_probe} to zero
intensity of the dressing beam gives the decay rate of the matter
wave grating $\Gamma_{2}$ of $(100 \, \mu {\rm s})^{-1}$, in fair
agreement with the linewidth of the Bragg excitation process
observed previously~\cite{sten99brag}. This observation of the
decay of the matter-wave grating can be regarded as pump-probe
spectroscopy of quasi-particles in the condensate.  The seeding
Bragg pulse created the quasi-particles (in this case condensate
excitations in the free-particle regime). One can control the
momentum of the excited quasi-particles by the angle between the
laser beams. This could be used to excite phonon-like
quasiparticles~\cite{stam99phon}, and their lifetimes could be
determined with the pump-probe scheme presented here.

\subsection{Single-atom and collective behavior}

\begin{figure}
    \begin{center}
    \epsfig{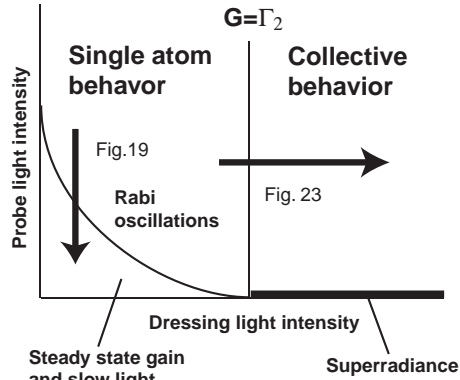}
    \caption{
     Schematic diagram of the different regimes of a dressed condensate. Depending on the
    intensities of the dressing and the probe beams, the dressed condensate occupies different
    physical regimes. Single atom and collective behavior are separated by the threshold to
    superradiance ($G=\Gamma_2$, Eq.\ \ref{eq:gain-equation}). }
    \label{fig:phasediagram}
    \end{center}
\end{figure}

The optical gain studied above clearly showed the transition from
single-atom gain (the first term in the expansion in
Eq.~\ref{eq:opt_gain_expansion}) to collective gain. Varying the
intensities of probe and dressing light allows for the study of
different physical regimes. At low dressing light intensity,
below the superradiant threshold, one encounters single-atom
behavior, at high intensity the system shows collective
superradiance.  The probe laser intensity determines whether the
system shows oscillatory or steady state response, as derived
above using optical Bloch equations.  Fig.\
\ref{fig:phasediagram} summarizes the different regimes.

\begin{figure}
    \begin{center}
    \epsfig{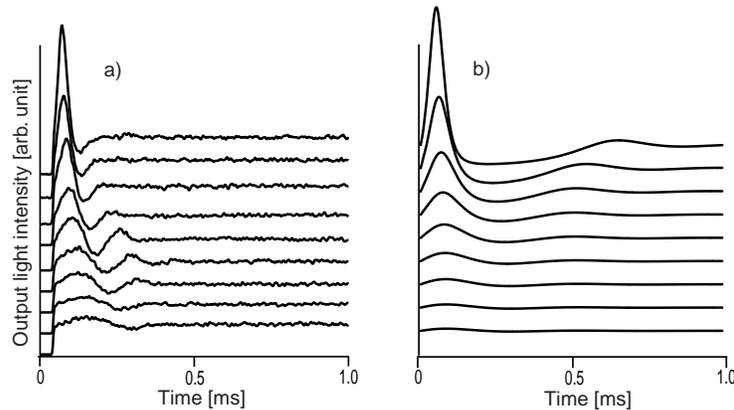}
    \caption{
    From single-atom Rabi oscillations to collective superradiance. a) The dressing beam intensity
    was increased from 2.7, 3.8, 6.1, 9.1, 14.1, 21.3, 32.7, 49.5 to 76 mW/cm$^2$ (bottom to top).
    The probe beam intensity was kept at 5 mW/cm$^2$. b) Numerical solution of the nonlinear
    optical Bloch equations (Eqs.\ \ref{omegadot}, \ref{vdot1} and \ref{wdot}).
    Plotted is $\Omega_R^2$ for the same experimental parameters as
in a).}
    \label{fig:Rabi_to_SR}
    \end{center}
\end{figure}

Probe light traces showing the transition from Rabi oscillations to superradiance are presented in
Fig.\ \ref{fig:Rabi_to_SR}.  As a function of the dressing light intensity, the damped Rabi
oscillations become faster and almost suddenly turn into a giant superradiant pulse.

Previously, recoil related gain based on single-atom phenomena
(Recoil Induced Resonances) was observed in cold cesium
atoms~\cite{cour94}. Collective gain due to the formation of a
density grating was discussed as a possible gain mechanism for
lasing action~\cite{boni94} (named CARL---Coherent Atomic Recoil
Laser) and pursued experimentally~\cite{lipp96CARL,hemm96} with
ambiguous results (see~\cite{brow97} and the discussion
in~\cite{moor99pra,berm99}). Our experiments clearly identify the
two regimes and their relationship.

The dressed condensate is a clean, model system for discussing optical and atom-optical properties.
The observed slow group velocity of the probe laser pulse can be directly related to the dynamics
of the amplification process.  The optical amplification can be described as a reflection of the
dressing light by a matter wave grating. The initial delay time in the amplification of optical
pulses is the time necessary to build up the (quasi-)steady state matter wave grating. When the
input pulse is switched off, the matter wave grating still exists and diffracts the pump light (as
observed in Fig.\ \ref{pump_probe}) creating the trailing edge of the transmitted light pulse.
Thus, the slow speed of light is simply related to the slow build-up and decay of quasi-particles
which we were able to monitor directly. In this microscopic picture, all photons propagate with the
vacuum speed of light $c$, the slow group velocity is only a phenomenological description of the
center-of-mass propagation of the amplified pulse. This description leads to the same number of
photons inside the condensate.  Slow light pulses are compressed by a factor $c/v_g$, but the
electric field strength is unchanged \cite{haus84,harr99prl}. Therefore, the product of the total
number of photons within the pulse and the transit time is constant.

\begin{figure}
    \begin{center}
    \epsfig{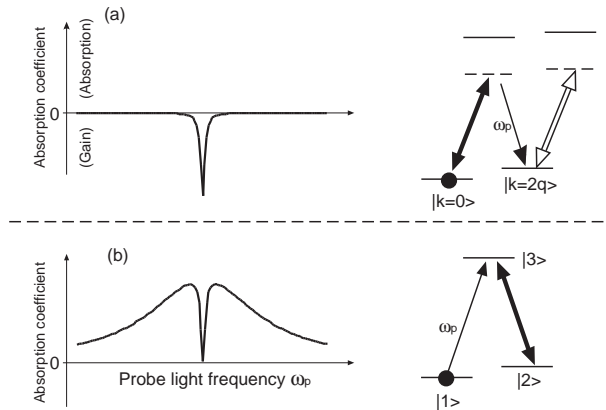}
    \caption{
    Comparison between two different methods to generate slow light. (a) Narrow-band optical
    amplification using a two-photon resonance.  (b) Electromagnetically induced transparency.}
    \label{fig:EIT}
    \end{center}
\end{figure}

Recent demonstrations of slow group velocities for light focused on electromagnetically induced
transparency (EIT) in a three-level $\Lambda$ system~\cite{hau99}. This system features a narrow
dip in a broad absorption feature. In our system, the broad absorption line is missing. Since the
propagation of resonant laser pulses is mainly determined by the narrow feature (which determines
${\rm d} n_1/{\rm d}\omega$), both systems show analogous behavior (see Fig.~\ref{fig:EIT}).
Indeed, if one would add a broad-band absorber to the narrow-band amplifier, one would create the
same index of refraction as in the EIT scheme.

Although both schemes involve three levels in a $\Lambda$
configuration (Fig.~\ref{fig:EIT}), there are major differences.
The amplification scheme does not have a dark state because it has
off-resonant couplings to other momentum states which are
indicated in Fig.~\ref{fig:EIT}. In the amplification scheme the
strong pump pulse connects the initially populated state to the
excited state in a far-off resonant way. In the EIT scheme the
strong coupling laser drives the other leg of the $\Lambda$
transition and is on resonance.

Finally, we want to resume the discussion of Sect.~\ref{sec:four_wave_mixing}, where we presented
the two limits of four-wave mixing of light and atoms which correspond to diffraction of light off
an atomic grating (for $N_q > n_{k-q}$) and diffraction of atoms off an optical grating (for $N_q <
n_{k-q}$).  These two limits of atomic stimulation and optical stimulation reflect whether the
final state occupation number is larger for atoms or photons. Therefore, we have to address what
are the relevant modes and their occupation numbers. So far, we have disregarded the absolute
number of photons.  The coupling constant $C$ (Eq.~\ref{eq:constant_C_phi}) is inversely
proportional to the mode volume $V$, hence for the physical description it was only relevant to
look at the photon density which is proportional to the electric field squared. Therefore, the
system can be well described semi-classically by the coupling of an electric field (or Rabi
frequency) to the atomic density, and on the previous pages, we have presented several formalisms
for that.

The question of the photon number becomes relevant when we want to use the concepts of
Sect.~\ref{sec:scattering} where we have developed a formalism to describe amplification in a BEC.
We showed that the scattering rate $W$ for a condensate illuminated by a laser beam is proportional
to
\begin{equation}
    W \propto N_0 n_k (N_q + n_{k-q} +1).
    \label{eq:W_general}
\end{equation}
This description focuses on two modes for the photons. However, since the experiment takes place in
open space with freely propagating laser beams, one has to carefully identify the relevant modes
for such a description.  Here we show that all the different cases of light scattering discussed in
these notes, Bragg scattering, Rayleigh scattering, and the amplification of light and atoms, can
be intuitively described by Eq.~\ref{eq:W_general}.

For Bragg scattering, the optical Bloch equations yield the steady state scattering rate as $W=N_0
\dot{w}/2=N_0 \Omega_R v/2=N_0 \Omega_R^2/\Gamma_2$.  This is in agreement with
Eq.~\ref{eq:total-stim-rate} when we replace the $\delta$ function by a (normalized) Lorentzian
lineshape of FWHM of $\Gamma_2$, which has a value of $2/\pi \Gamma_2$ on resonance. Thus the rate
for Bragg scattering can be written as (neglecting numerical factors and setting $S(q)=1$ for large
momentum transfer)
\begin{equation}
    W_{\rm Bragg} \approx (|C|/\hbar)^2 N_0 n_k (n_{k-q}+1)/ \Gamma_2.
    \label{eq:W_bragg}
\end{equation}

For the amplification of atoms, the gain equation (Eq.~\ref{eq:gain-equation}) implies
\begin{equation}
W_{\rm atom\_gain}  \approx (|C|^2/\hbar) N_0 n_k (N_q +1) \frac{{\rm d}\rho}{{\rm d}E {\rm
d}{\Omega}} \Omega_q.
\end{equation}
For the amplification of light, the scattering rate at low intensity of the dressing beam is given
by Eq.~\ref{eq:W_bragg}---thus we are in the regime of scattering atoms off an optical grating. The
number of photons $n_{k-q}$ in the input mode is the number of photons in a volume $V$ with the
cross section $A$ of the condensate and a length $l_{\rm mode} \approx c/ \Gamma_2$.  This choice
of length is the coherence length of light with a spectral bandwidth $\Gamma_2$. This value for
$l_{\rm mode}$ can be justified by another argument.  The coupled equations
(Eq.~\ref{eq:coupled_eq_damp}) reduce the physical description to one mode of the scattered light.
The angular mode selection occurs by the cigar shape of the condensate.  The Bragg scattering
process with a linewidth $\Gamma_2$ couples to only one longitudinal mode if we assume boundary
conditions which impose a mode spacing of $\Gamma_2$, e.g.\ assuming non-reflecting walls at a
distance $\approx c/ \Gamma_2$. To be consistent, the number of photons in the input laser beam has
to be determined over the same length.

Now we want to turn to the transition from light amplification stimulated by light to light
amplification stimulated by atoms.  Using Eq.~\ref{eq:Rabi_and_Rayleigh} one obtains (again
neglecting numerical factors)
\begin{equation}
    n_{k-q} \approx \Omega_R^2 A l_{\rm mode} /R \lambda^2 c .
\end{equation}
Expressing the Rayleigh rate $R$ by the atom gain $G$ (Eq.~\ref{eq:gain1}), $R \approx G A/N_0
\lambda^2$, we get
\begin{equation}
    n_{k-q}  \approx N_0 \Omega_R^2 l_{\rm mode} /G c \approx N_0 \Omega_R^2 /G \Gamma_2.
    \label{eq:n_k}
\end{equation}
The number $N_q$ of recoiling atoms is approximately the Bragg scattering rate,
$\Omega_R^2/\Gamma_2$ times the coherence time $1/\Gamma_2$ times the number of condensate atoms
$N_0$
\begin{equation}
    N_q \approx N_0 \Omega_R^2/\Gamma_2^2. \label{eq:N_q}
\end{equation}
Comparing Eqs.~\ref{eq:n_k} and \ref{eq:N_q} we see that around the threshold for superradiance
($G=\Gamma_2$), the regime $N_q \ll n_{k-q}$ crosses over to $N_q \approx n_{k-q}$.  In other
words, the bosonic stimulation factor during the optical amplification, $(N_q + n_{k-q} + 1)$
(Eq.~\ref{eq:W_general}), is approximately $n_{k-q}$ in the linear (single-atom) regime. Near the
threshold to superradiance, the $N_q$ term becomes significant and diverges at threshold. This
statement can be generalized to include superradiance.  Whenever the number of recoiling atoms
$N_q$ dominates in Eq.~\ref{eq:W_general} one enters the regime of nonlinear amplification. Without
a probe beam, i.e.\ for $n_{k-q}=0$, nonlinear scattering and superradiance set in for $N_q \approx
1$.

The photon counting becomes even more confusing when one tries to extend it to the scattered
photons. The following remarks are rather tentative and reflect hours of controversial and not
fully resolved discussions. The occupation number of the adiabatically eliminated photon field in
Eq.~\ref{eq:ad_elim} is $n_{k-q} = |c^*|^2 \approx G  N_q/ \Gamma_c$ (with $N_q=|a|^2$). Since the
gain $G$ is the Rayleigh scattering rate into the solid angle $\Omega_q$ of the mode, we can
interpret $n_{k-q}$ as the photon number emitted into the mode during the damping time $1/\Gamma_c$
of the photon field (and stimulated by $N_q$ atoms in the final state). The physical process is
independent of the damping time $\Gamma_c$ (and therefore of the absolute number $n_{k-q}$ of
photons). In our experiments, the scattered photons hit the walls of the vacuum chamber (or a photo
detector) within nanoseconds.

In the case of atom amplification, all scattered photons were emitted spontaneously (in the sense
of the superradiant cascade described by Dicke \cite{dick54}) and should not be counted in a
bosonic stimulation term. Even if these photons were emitted into a cavity, they would stimulate
the emitting system only after being reflected back. However, it is precisely this photon field
(Eq.~\ref{eq:ad_elim}) which leads to the bosonic stimulation by atoms (Eq.~\ref{eq:single_eq}).
The Hamiltonian (Eq.~\ref{eq:simple_ham}) couples atoms only to light.  After eliminating the
photon field, the (atomic) bosonic stimulation term represents the coupling of all atoms to the
same mode of the electromagnetic field, and that's how the atoms stimulate each other. Rate
equations (Eq.~\ref{eq:W_general}) only apply after coherences have damped out.  The most rapidly
damped mode is therefore no longer an independent degree of freedom, but slaved to the slower
modes. This suggests that the photon number $n_{k-q}$ in Eq.~\ref{eq:W_general} reflects the
external laser field, and not the scattered photons---they appear already in the atomic stimulation
term $N_q$. All these conceptional problems are avoided (but not solved!) when one sticks to the
coupled equations between the photon and atom fields.

If we inject probe light near the threshold to superradiance, we have optical amplification by
bosonic stimulation of atoms! However, the output light intensity is still proportional to the
input intensity.  In steady state, the population $N_q$, the Rayleigh scattering rate (stimulated
by these $N_q$ atoms) and therefore the number of photons added to the probe light are all
proportional to the input light intensity.  The system is driven by the probe light input.  It is
amplified by a gain factor which includes the dynamics of the coupled atomic and optical fields.

Finally, we want to show how to obtain the Rayleigh scattering rate from Eq.~\ref{eq:W_bragg}.  We
assume the final photon states to be empty (setting $n_{k-q}=0$) and multiply with the number of
final states. This number $\Delta \rho$ is obtained by multiplying the density of states
Eq.~\ref{density_of_states} by the linewidth $\hbar \Gamma_2$ and the full solid angle $\Delta
\rho= ({\rm d}\rho/{\rm d}E {\rm d}{\Omega}) ( 4 \pi)(\hbar \Gamma_2)$. This leads to
\begin{equation}
    W_{\rm Rayleigh} = (|C|/\hbar)^2 N_0 n_k  \Delta \rho /\Gamma_2
    \approx (|C|^2/\hbar) N_0 n_k  \frac{{\rm d}\rho}{{\rm d}E {\rm d}{\Omega}} 4 \pi,
\end{equation}
which is approximately equal to the Rayleigh scattering rate $R$
(Eq.~\ref{eq:scatt_rate_solid_angle}).  The stimulated Bragg scattering rate (Eq.~\ref{eq:W_bragg})
dominates over the spontaneous Rayleigh rate when the number of photons $n_{k-q}$ in the Bragg beam
is larger than the number $\Delta \rho$ of accessible modes. Note that this intuitive result is
independent of defining modes and boundary conditions---the two-photon linewidth $\Gamma_2$ has
cancelled out. Subtleties arise only in the case of coherent emission. In this case, one has either
to describe coherent superposition states of several modes, or identify modes which reflect the
linewidth of the transition.

\section{Enhancement of spontaneous emission in a Bose-Einstein condensate}
\label{sec:enhanced_spont_em}

In the previous sections we have illustrated the rich physics
which is described by the rate equations introduced in the first
section. They suggested that the relevant matrix element is
always $S(q)$, and that enhanced scattering relies on the
population of the final states with several particles or photons.
This and the following section broaden this picture. This section,
which is a slightly modified version of a recently submitted
paper \cite{gorl00spont}, shows that $S(q)$ is the matrix element
only for scattering particles or light, but not for spontaneous
emission.  Finally, Sect.\ \ref{sec:MWA_fermions} discusses that
collective enhancement is not restricted to bosonic systems and
macroscopic population of a quantum state.

In the previous sections we have shown that light scattering and particle scattering in a
condensate can be suppressed.  This suppression is due to the reduction of density fluctuations in
a condensate at long wavelengths (longer than the healing length).  The dissipation-fluctuation
theorem implies that the same suppression can be observed in the response of a condensate to an
external perturbation. Therefore, one might interpret the suppression as the ``unwillingness'' of a
condensate against long-wavelength modulations.

A similar suppression of the long wavelength response (or reduction of the structure factor
$S(q)$) occurs in degenerate Fermi gases due to Pauli blocking.  Transfer of momentum comparable to
or less than the Fermi momentum is suppressed due to occupancy of possible final states. Pauli
blocking is a single-particle effect and would occur even in an ideal non-interacting Fermi gas.
In contrast, the suppression of momentum transfer in a BEC is a genuine many-body effect due to the
atomic interactions and the collective nature of the condensate. It seems that Fermi seas and
condensates have similar behavior for different physical reasons.

In addition to particle and light scattering, we want to discuss now the process of spontaneous
emission of an excited atom initially at rest in the condensate or the Fermi sea.  For the Fermi
sea, there is suppression:  spontaneous emission is impossible for recoil momenta less than the
Fermi momentum. Surprisingly, the situation is different for a condensate where spontaneous
emission is enhanced.

This effect is most easily derived for a homogeneous condensate
at zero temperature consisting of $N$ atoms at a density $\rho$
in a volume $V$. Due to the atomic interactions described by
Eq.~\ref{eq:hamil_indist}, two atoms in the zero-momentum state
are coupled to states with momenta $+\hbar{\bf k}$ and $-\hbar{\bf
k}$. The ground state wavefunction $|g \rangle$ of a BEC has
admixtures of pair correlations yielding the structure
\cite{huan87}
\begin{equation}
    \label{equ:bec_ground_state}
    |g \rangle = |N,\,0,\,0\rangle -
    \alpha |N-2,\, 1,\,1\rangle + \alpha^2 |N-4,\, 2,\, 2 \rangle + \dots ,
\end{equation}
where $\alpha=1-1/u_k^2$. Here $|N_0, N_k, N_{-k}\rangle$ denotes
a state with $N_0$ atoms in the zero-momentum state and $N_{\pm
k}$ atoms in states with momentum $\pm \hbar k$. In
Eq.~\ref{equ:bec_ground_state} a summation over all momenta
$\hbar {\bf k}$ is implicitly assumed. The average population of
momentum states is given by $N(k)= u_{k}^{2}-1 = v_{k}^{2}$,
where $u_k = \cosh \phi_k$, $v_k = \sinh \phi_k$ and $\tanh
2\phi_k = \mu / (\hbar \omega_k^0 + \mu)$ as already introduced
in Sect.\ \ref{sec:Bragg_spect}.

To study the effect of the presence of a BEC on spontaneous
emission, we consider an excited atom at rest added to a BEC of
$N$ ground-state atoms. This system is described by an initial
state $|i\rangle = \hat{a}_{e,0}^\dagger |g \rangle$, where
$\hat{a}_{e,0}^\dagger$ creates an electronically excited atom at
rest. We use Fermi's golden rule to obtain the rate for
spontaneous emission.  The only difference to the single-atom
spontaneous decay rate $\Gamma$ comes from the overlap matrix
elements to the final momentum state $|f\rangle$, $\langle
f|\hat{a}_{k_L}^\dagger \hat{a}_{e,0} | i \rangle$ where
$\hat{a}_{k_L}^\dagger$ is the creation operator for a free
ground state atom with momentum of the photon $\hbar {\bf k_L}$.
Summing over all final states one arrives at
\begin{eqnarray}
    \label{equ:spont_BEC}
    \gamma_{\text{BEC}}=\Gamma \,   \langle g |\hat{a}_{k_L}
    \hat{a}_{k_L}^\dagger|g \rangle \, .
\end{eqnarray}
Thus, the spontaneous emission rate is proportional to the square of the norm of the state vector
$|e^+ \rangle = \hat{a}_{k_L}^\dagger|g \rangle $.

To calculate the norm of $|e^+ \rangle$ explicitly, we transform to Bogoliubov operators (see
Sect.\ \ref{sec:Bragg_spect}) and obtain
\begin{eqnarray}
    \label{equ:f_bose_spont}
    F_{\text{Bose}}^{\text{spont}} &=& \langle e^+|e^+ \rangle = u_{k_L}^2
    =  1+ N(k_L)  \\ \nonumber
    &=& \left( \cosh \,\left( \frac{1}{2}\, \tanh^{-1}\left(\frac{k_s^2}{k_L^2/2 +
    k_s^2}\right)\right) \right)^2 \, ,
\end{eqnarray}
where $\hbar k_s = M c_s$ is the momentum of an atom moving at the speed of sound. Enhancement of
spontaneous emission in a BEC is significant if $k_s$ becomes comparable to the wavevector $k_L$ of
the emitted photon since for small momentum transfer $u_{k_L}^{2}= k_s^2/k_L^2$.

For comparison, we briefly summarize the suppression of spontaneous emission and light scattering
for a fermionic system. A Fermi gas at $T = 0$ with Fermi momentum $\hbar k_F$ is characterized by
$N({\bf k}) = \theta(k_F-k)$, i.e.,\ all momentum states with $k < k_F= (6 \pi \rho)^{1/3}$ are
occupied. If we add an electronically excited atom at rest to the Fermi sea, its spontaneous decay
rate is suppressed by a factor
\begin{equation}
    F^{\text{spont}}_{\text{Fermi}} = 1-N(k_{L})=\theta(k_{L}-k_F)\, .
    \label{equ:f_fermi_spont}
\end{equation}

\begin{figure}
    \epsfxsize=55mm
    \centerline{\epsfbox{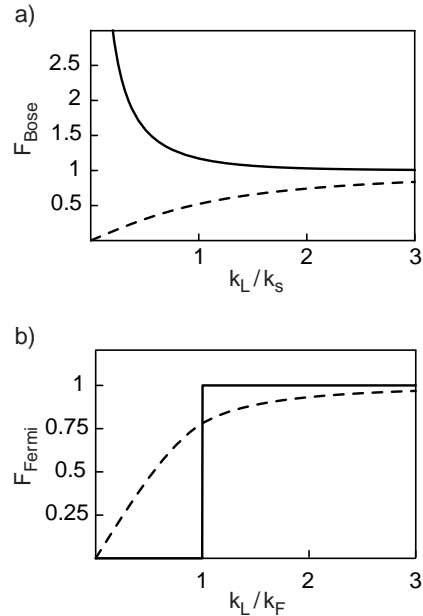}}
    \caption{Modification of spontaneous emission (solid line) and
    light scattering (dashed line) due to quantum degeneracy. In a) we have plotted the enhancement
    factor for spontaneous emission and the suppression factor for light scattering for a weakly
    interacting Bose-Einstein condensate as a function of the light wavevector $k_L$ in units of
    $k_s$, the wavevector of an atom moving at the speed of sound. In b) the suppression factors for
    spontaneous emission and light scattering in a Fermi gas at $T=0$ are plotted as a function of
    $k_L$ in units of the Fermi wavevector $k_F$.  Figure is taken from Ref.\ \protect\cite{gorl00spont}.}
    \label{spont_em}
\end{figure}

When off-resonant light with initial wavevector ${\bf k_L}$ is scattered from a filled Fermi
sphere into an outgoing wave with final wavevector ${\bf k_L}+{\bf q}$, the scattering rate is
suppressed by \cite{pine88}
\begin{eqnarray}
    \label{equ:fermi_structure_factor}
     S_{\text{Fermi}}({\bf q}) &=& \int {\rm d} {\bf k}\ N({\bf k})
    (1-N({\bf k+q}))\\ \nonumber &=& \cases {\frac{3 q}{4 k_F} - \frac{q^3}{16 k_F^3} & \text{if $0 < q
    <2 k_F$}, \cr\cr 1 & \text{if $q >2 k_F$}.}
\end{eqnarray}
Eq.\ \ref{equ:fermi_structure_factor} is the static structure factor for a Fermi gas at zero
temperature. Integrating over all possible scattering angles and accounting for the dipolar
emission pattern, we find that the total suppression factor for Rayleigh scattering from a Fermi
sea is given by
\begin{equation}
    \label{equ:f_fermi_scatt}
     F^{\text{scatt}}_{\text{Fermi}} = \cases {\frac{69}{70} \frac{k_L}{k_F}
    - \frac{43}{210} \frac{k_L^3}{k_F^3} & \text{if $k_L <k_F$} \cr\cr 1 - \frac{3}{10}
    \frac{k_F^2}{k_L^2} \, + \, \frac{9}{70} \frac{k_F^4}{k_L^4}  - \frac{1}{21} \frac{k_F^6}{k_L^6} &
    \text{if $k_L>k_F$}.}
\end{equation}

Fig.\,\ref{spont_em} shows the influence of quantum degeneracy on the atom-light interaction. Using
Eqs.\ \ref{equ:f_bose_spont}, \ref{equ:f_bose_scatt}, \ref{equ:f_fermi_spont}, and
\ref{equ:f_fermi_scatt} we have plotted the rates for spontaneous emission (solid lines) and light
scattering (dashed lines), normalized by the single-atom rates, for a weakly interacting BEC
(Fig.\,\ref{spont_em}a) and a degenerate Fermi gas (Fig.\,\ref{spont_em}b). A significant deviation
from the free-particle rate is clearly observable if the photon-momentum is comparable to $k_s$ for
bosons and $k_F$ for fermions.

The enhancement of spontaneous emission in a BEC can be
intuitively explained as bosonic stimulation by the quantum
depletion since the enhancement factor ($1+ N(k_L)$) has the same
form as if $N(k_L)$ atoms would occupy the final state. This
intuitive argument is correct, but it would incorrectly predict
that light {\it scattering} is also enhanced in contrast to what
we have shown in Sect.\ \ref{sec:Bragg_spect}. The suppression of
light scattering occurs due to the correlation between the
admixtures of states with momentum $\hbar {\bf k}$ and $-\hbar
{\bf k}$. This leads to a destructive quantum interference between
the two processes $|N,0,0\rangle + \hbar q \rightarrow |N,1,0
\rangle$ and $|N-2,1,1\rangle + \hbar q \rightarrow |N,1,0
\rangle$, in which either an excitation with momentum $\hbar
\bf{q}$ is created
 or an excitation with momentum $- \hbar \bf{q}$ is annihilated. Both processes transfer momentum $\hbar \bf{q}$
to the condensate and are individually enhanced by bosonic stimulation.  Therefore, a simple rate
equation model would predict enhanced light scattering. However, since the initial states are
correlated the two processes leading to the same final state interfere destructively for a BEC
with repulsive interactions and light scattering is suppressed.

The static structure factor $S(q)$ of a condensate and the reduced light scattering can be obtained
from the Gross-Pitaevskii equation as the response to a periodic perturbation
\cite{stamp00leshouches}. However, the enhanced spontaneous emission is related to the injection
of an additional atom into a condensate and appears to be physics beyond the Gross-Pitaevskii
equation.

How strong would the enhancement of spontaneous emission in
currently realized Bose-Einstein condensates be? Condensates of
$^{23}\text{Na}$ atoms confined in an optical trap  have reached a
density of $3 \times 10^{15} \, {\rm cm}^{-3}$ \cite{stam98odt}.
For this density the speed of sound $\hbar k_s/M = 2.8 \rm \,
cm/s$ and the recoil velocity $\hbar k_L/M = 2.9 \rm \, cm/s$ are
approximately equal and we find $N(k_L) \approx 0.15$. Thus, the
observation of enhanced spontaneous emission in a BEC is within
experimental reach.  Excited atoms at rest could be produced by
injecting ground-state atoms with momentum $\hbar \bf{k_L}$ into a
condensate and using a counter-propagating laser beam to excite
them and bring them to rest.  The enhancement of spontaneous
emission could then be observed as frequency broadening of the
absorption line.

The fact that light scattering is suppressed, but spontaneous emission is enhanced, could be
exploited for studies of decoherence in a BEC.  When a photon is absorbed by a BEC (the first step
of light scattering), it creates a (virtual) excited state that has an external wavefunction which
includes pair correlations.  Any decoherence of this coherent superposition state, for example by
interaction with the thermal cloud, could destroy the interference effect discussed above and turn
the suppression of light scattering into an enhancement. Another possibility of creating an
excited state atom in a BEC is using Doppler free two-photon excitation, a scheme already used to
probe condensates of atomic hydrogen on the $1s \rightarrow 2s$ transition \cite{frie98}. In this
case, enhancement of spontaneous emission could be observed if the excited state lifetime is
longer than the coherence time.

\section{Does matter wave amplification work for fermions?}
\label{sec:MWA_fermions}

{\it Introduction}. In the previous sections we have discussed several examples of bosonic
stimulation for massive particles, bosonically enhanced elastic collisions, superradiance of atoms
\cite{inou99super}, and matter wave amplification \cite{inou99mwa,kozu99amp}. This and the four
wave mixing of atoms \cite{deng99} were described as processes which are bosonically stimulated,
i.e.,\ their rates are proportional to $(N_q+1)$, where $N_q$ is the number of identical bosons in
the final state. These experimental achievements have raised the question whether these processes
are inherently connected to bosonic systems.

At the Cargese summer school, we presented the view that all these processes do not depend on
Bose-Einstein statistics and would occur for thermal atoms or even for fermions, although with a
much shorter coherence time \cite{inou99super}.  These suggestions have stirred many controversial
discussions at the summer school.  This section will reconcile the different physical descriptions.
The central result is that the stimulated processes mentioned above do not rely on quantum
statistics, but rather on symmetry and coherence.  This section is a slightly extended version of a
recently submitted paper \cite{kett00fermi}.

We also address a widespread misconception about bosonic stimulation which regards stimulated
scattering as being solely due to quantum-statistical enhancement by the final state, i.e.,\ as if
the particles in the final state attract other identical particles {\it without any other physical
effect}.  We show that the presence of a macroscopically occupied state increases the density
fluctuations of the system, and bosonically enhanced scattering is simply the diffraction of
particles from these density fluctuations.  First, we establish the equivalence of bosonically
enhanced scattering, diffraction and superradiance which will then be applied to fermionic systems.

{\it Scattering theory}. Sect.\ \ref{sec:scattering} presented
basic aspects of the theory of scattering of light or particles
from an arbitrary system.  These results simply followed from
lowest order perturbation theory (Fermi's Golden Rule). The
double differential cross-section for scattering can be
decomposed into two factors $\frac{{\rm d}^2\sigma}{{\rm d}\Omega
\: {\rm d}\omega}= \left( \frac{{\rm d}\sigma}{{\rm d}\Omega}
\right)_{\rm single} S(q,\omega)$.  The first one is the
differential cross-section for the scattering by a single
particle (e.g. the Rayleigh cross-section for far-off resonant
light scattering), the second one is the dynamic structure factor
(van Hove or scattering function) $S(q, \omega)$ which is the
Fourier transform of the density-density correlation function: $
S(q,\omega)=(1/2\pi) \int{{\rm d}t \: e^{{\rm i}\omega t} \langle
\hat\rho(q,t) \hat\rho^\dag(q,0)} \rangle $ where $\hat\rho(q)$
is the Fourier transform of the particle density operator
introduced in Sect.\ \ref{sec:scattering}.

For a non-interacting system of bosons, $S(q,\omega)$ can be expressed using the single-particle
states $|i\rangle$ with energy $E_i$ and occupation numbers $N_i$ as
\begin{eqnarray}
    \lefteqn{ S(q,\omega ) = S_0(q) \delta(\omega) +}  \nonumber \hspace{0.15 in} \\
    && \sum\limits_{i \neq j}  \left| {\left\langle j \right|e^{{\rm i}qr} \left| i \right\rangle }
    \right|^2 N_i(N_j+1) \delta \left[ {\omega  - (E_j - E_i)/\hbar} \right].
    \label{eq:sqw}
\end{eqnarray}
The factor $(N_j+1)$ reflects bosonic stimulation by the occupation of the final state. We have
split off the elastic term $S_0(q)$ which describes coherent elastic scattering or diffraction and
is simply the square of the Fourier transform of the density $ S_0(q) = \left| \langle
\rho^{\dag}(q) \rangle \right|^2 =\left|  \sum  N_i {\left\langle i \right|e^{{\rm i}qr} \left| i
\right\rangle }\right|^2$.

{\it A simple example}. It is instructive to apply this formalism to a system of non-interacting
bosons which has macroscopic occupation in two momentum states with momentum $\pm \hbar \bf{k}$. If
the initial state is a Fock state $|+k \rangle^{N_+} |-k \rangle^{N_-}$, we find that, apart from
forward scattering, the dominant term in $S(q, \omega)$ is the bosonically enhanced scattering
between those two (degenerate) states, $S(q,\omega)= \left[ N^2 \delta_{q,0}+ N_+(N_-+1)
\delta_{q,-2k} + N_-(N_+ +1) \delta_{q,2k} \right] \delta(\omega)$ where the Kronecker symbol
$\delta_{q,p}$ implies ${\bf q}={\bf p}$ within the wavevector resolution $\approx 1/l$ of a finite
volume with length $l$.  Alternatively, we can assume the initial state to be a coherent
superposition state $|i \rangle^N$ with the eigenstate $|i \rangle= c_+ |+k \rangle + c_- |-k
\rangle$ and $|c_{\pm}|^2=N_{\pm}/N$ and $N=N_+ + N_-$. Now, the dominant contribution to
$S(q,\omega)$ comes from $S_0(q)=  N^2 \delta_{q,0} + N^2 |c_+| ^2 |c_-|^2 \left[ \delta_{q,2k} +
\delta_{q,-2k} \right]$ which is equivalent to the Fock state case when the difference between
$N_{\pm}$ and $N_{\pm}+1$ can be neglected in the limit of large occupation numbers.

This equivalence between Fock states and coherent superposition states has been extensively
discussed in the context of two interfering Bose-Einstein condensates
\cite{java96phas,nara96,cast97} and also with regard to optical coherences \cite{molm97}.  Those
papers show that, in many situations, a Fock state is equivalent to an ensemble of coherent states
with arbitrary phase. Experimental interrogation determines the phase and reduces the ensemble to a
single coherent state with a phase which will vary from experiment to experiment. For large
occupation numbers, one can therefore regard the Fock state as an initial state which has not yet
``declared its phase'', and, in many cases, for the convenience of calculations, replace the Fock
state by a coherent superposition state with an arbitrarily chosen phase.

However, on first sight, the physical interpretation is different. In the Fock state formulation,
the enhanced scattering results from a macroscopic occupation number in a single quantum state,
whereas for the coherent superposition state, the scattering is simple diffraction by a
sinusoidally modulated density distribution with an amplitude proportional to $N|c_+ c_-|$. This
density modulation acts as a diffraction grating for incident light or particles and has a
diffraction efficiency proportional to the square of the amplitude. Such a density modulation does
not require bosonic atoms.  It can, for example, be imprinted into thermal or fermionic clouds by
subjecting them to a suitable optical standing wave. The equivalence of these two descriptions
points towards one of the major conclusions of this section, namely that macroscopic population of
bosonic states is not necessary for enhanced scattering.

The previous discussion assumed scattering between two degenerate
momentum states $|\pm k \rangle$. A simple Gallilean
transformation generalizes this to two arbitrary momentum states
$|k_{\pm} \rangle$ with energies $E_{\pm}$.  Now the standing
wave moves with a velocity $\hbar (k_+ + k_- )/2 M$ where $M$ is
the mass of the atoms, and the enhanced scattering appears at
$\hbar \omega=\pm(E_+ - E_-)$ instead of at $\omega=0$.

{\it Enhancement of fluctuations}. The general results of statistical physics presented above
emphasize that enhanced scattering {\it must} be related to enhanced density fluctuations.
Therefore, bosonic enhancement of a scattering rate is either due to a density modulation $\langle
\rho(q) \rangle$ (in the coherent superposition description) or due to density fluctuations (in the
Fock state description)---the latter can be regarded as a density modulation with an unknown phase.
This relation allows a more intuitive answer to the question why is there bosonic enhancement when
two atoms 1 and 2 collide in the presence of a condensate with $N_0$ atoms.  The standard answer
would be that the symmetry of the wavefunction enhances the scattering rate into the condensate and
into some other state 3 by a factor of $(N_0 +1)$.  An equivalent answer is that the condensate
interferes with atom 2 (or 1) and creates a density grating with an amplitude proportional to
$N_0^{1/2}$ which diffracts atom 1 (or 2) into state 3. The grating absorbs this momentum transfer
by transferring the atom in state 2 (or 1) into the condensate. Therefore, bosonic stimulation can
be regarded as heterodyne amplification of density fluctuations where the condensate acts as the
local oscillator.  This alternative physical picture emphasizes the role of interference in bosonic
stimulation.

{\it Dicke superradiance}. We now want to establish the connection between bosonic enhancement and
Dicke superradiance.  This will formally introduce the enhancement factor $(N+1)$ for non-bosonic
systems. A system with $N$ atoms in two states $|\pm \rangle$ is conveniently described with the
formalism introduced by Dicke to discuss superradiance in two-level atoms \cite{dick54}. It should
be emphasized that the only assumption in this treatment is that the $N$ atoms couple identically
to the probe field (the electromagnetic field or some incident particle beam), i.e.,\ that they
have the same transition frequency and matrix element without any assumption of quantum
statistics.  For example, in magnetic resonance experiments, the Dicke treatment would apply to
different atomic species with the same value of the magnetic moment.

It should be emphasized that Dicke superradiance depends only on
the symmetry of the emitting system and does not depend on the
nature of the emitted particle, whether they are bosons or
fermions.  For example, the Dicke treatment would apply to an
ensemble of atoms in an autoionizing state which emit electrons.
If the ensemble is localized within a de Broglie wavelength of the
electron, enhanced superradiant emission would occur.  The
fermionic nature of the emitted particles is irrelevant in the
so-called microscopic regime \cite{gros82} where less than one
emitted particle is in the volume of the sample.  This is the
situation originally discussed by Dicke. Many experiments on
superradiant emission of light have been done in regimes with many
photons per mode which of course would not be possible with
fermions.

Dicke treated the two-level atom as a spin 1/2 system and introduced angular momentum quantum
numbers. In this subspace, a fully symmetric state of $N$ atoms has spin $s=N/2$ and magnetic
quantum number $m=(N_+ -N_-)/2$.  The squared matrix element for the transition $|s, m\pm 1 \rangle
\rightarrow |s, m \rangle$ induced by the ladder operator $S_\mp$ is $(s \pm m +1)(s \mp m)$.
Expressing this by initial occupation numbers $N_\pm$, one obtains $N_\pm (N_\mp +1)$
\cite{sarg74,saku94,wall97mwa} retrieving the formula of bosonic enhancement. The transition rates
are largest for the $N$ particle state with $s=N/2$ which is therefore called the state of maximum
cooperativity.

Such a system will couple to the probe field in a superradiant way (i.e.\, with an up to $N$ times
enhanced transition rate).  In the Bloch vector picture, its dynamics is described as the
precession of a macroscopic spin vector with length $s=N/2$. This spin vector decays in a time
$1/\Gamma$ where $\Gamma$ is the total (homogeneous and inhomogeneous) linewidth of the transition
$|+ \rangle \rightarrow |- \rangle$. Collective superradiant behaviour can only be observed at
times shorter than $1/\Gamma$.

\begin{figure}[htbf]
    \epsfxsize=84 mm \centerline{\epsfbox{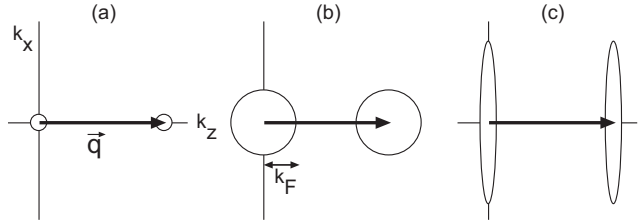}}\vspace{0.3cm}
    \caption{Momentum transfer ${\bf q}$ (a) to a Bose-Einstein condensate, (b) to a Fermi
    sea, and (c) to a momentum squeezed degenerate Fermi cloud.  Shown are the populated states
    vs. the $k$-vector. The momentum spread $k_F$ of the Fermi sea introduces Doppler broadening
    of the transition and a finite coherence time, whereas the coherence time in (a) and (c) is
    infinite. Figure is taken from Ref.\ \protect\cite{kett00fermi}.}
    \label{fig:Momentum_transfer}
\end{figure}

{\it Matter wave gratings and fermions}. Dicke's formalism is usually applied to one-photon
transitions between internal states, but here we use it to discuss scattering, i.e.,\ a two-photon
transition between two momentum states $|k_{\pm} \rangle$.  Let's first assume that we have an
ideal Bose-Einstein condensate in the ${\bf k}=0$ momentum state. Light scattering between momentum
states ${\bf k}=0$ and ${\bf k}={\bf q}$ has an infinite coherence time for a non-interacting
condensate of infinite size (Fig.~\ref{fig:Momentum_transfer}a). For a thermal (non-degenerate)
cloud of atoms with thermal momentum spread $\hbar k_{\rm th} \ll \hbar q$ the transition for the
transfer of momentum $\hbar q$ is Doppler broadened by $\Gamma=\hbar k_{\rm th} q/m$. For times
shorter than $1/\Gamma$ the system will behave collectively like the Bose condensed system, i.e.,\
a probe beam would induce transitions between the ${\bf k}=0$ and ${\bf k}={\bf q}$ momentum states
at a rate proportional to $N_{k=0} (N_{k=q} + 1)$ where $N_{k=0,q}$ refers to the total number of
atoms in states with momentum around ${\bf k}=0, {\bf q}$.

Once we have distributed the particles over many initial states,
indistinguishability and quantum statistics don't play any role.
Therefore, the only modification for a Fermi degenerate cloud is
to replace $k_{\rm th}$ with the Fermi wavevector $k_F$ in the
expression for the inhomogeneous broadening
(Fig.~\ref{fig:Momentum_transfer}b). Due to the assumption $\hbar
k_F \ll \hbar q$, Pauli blocking due to scattering into already
occupied states is absent. If this assumption is not made, a part
of the cloud becomes inactive, and our discussion would apply
only to the atoms near the Fermi surface.

The previous paragraph generalized the bosonic {\it Fock state}
ensemble to non-degenerate and fermionic clouds.  We now come
back to the {\it coherent superposition} state. For bosons, it can
be produced from a Bose-Einstein condensate in the ${\bf k}=0$
state by applying a (so-called Bragg) pulse of two laser beams
which differ in wavevector by ${\bf q}$ and in frequency by the
recoil frequency $\hbar q^2/2m$. Those beams resonantly drive the
transition between momentum states ${\bf k}=0$ and ${\bf k}={\bf
q}$ \cite{kozu99bragg,sten99brag} and prepare the superposition
state discussed above. Similarly, in a thermal (or fermionic)
cloud, the Bragg pulse creates a modulated density distribution
with wavelength $2 \pi/q$ which has the same contrast as in the
bosonic case and will diffract light or atoms at the same rate.
However, due to the thermal motion with velocity $\hbar k_{\rm
th}/M$, this grating decays during a time $M/\hbar k_{\rm th} q =
1/ \Gamma$ (for the fermionic case, $k_F$ has to be substituted
for $k_{\rm th}$). Thus the Dicke picture and the diffraction
picture agree.

{\it Coherence time}. The Doppler broadening discussed above
seems to imply a fundamental limit to the coherence time of a
Fermi system.  However, at least in principle, one can prepare a
Fermi system with infinite coherence time by starting out with a
cloud which is in a single momentum state along the $\hat z$
axis, but occupies many momentum states along $\hat x$ and $\hat
y$. With a Bragg pulse transferring momentum $\hbar q \hat z$, one
can prepare a system which shows collective behavior for
scattering particles or light with momentum transfer $\hbar q \hat
z$ with an infinite coherence time
(Fig.~\ref{fig:Momentum_transfer}c). Therefore, there is no direct
connection between a long coherence time and a high phase-space
density. In this ensemble, the scattering is between the states
$|k_z=0 \rangle \otimes |k_x, k_y \rangle$ and $|k_z=q \rangle
\otimes |k_x, k_y \rangle$. Therefore, we have enhanced
scattering into the $|k_z=q \rangle$ quantum state, but the atoms
may differ in other quantum numbers. What matters is only the
symmetrization of the many-body wavefunction along $\hat z$.  The
other quantum numbers ensure that there is no conflict with the
Pauli blocking for fermionic systems.  This is analogous to the
separation of electronic wavefunctions into a symmetric part
(e.g. the spin part) and an antisymmetric part (e.g. the spatial
part) where the coupling to an external field (e.g. electron spin
resonance experiment) only depends on the symmetric part.

{\it Experiments}. The experiments both on superradiance
(Sect.~\ref{sec:SR_and_MWA} and \cite{inou99super}) and four-wave
mixing \cite{deng99} in Bose-Einstein condensates have in common
that a matter wave grating formed by two macroscopically occupied
momentum states is probed, either by light or by atoms. Both
experiments create the coherent superposition state discussed
above using a Bragg pulse. In the limit of low intensity of the
probe beam, the scattering is independent of the nature of the
probe particles---one could have used any kind of radiation,
bosons or fermions \cite{vill00fwm}. The bosonic stimulation
observed in both experiments demonstrates the dynamic nature of
the matter wave grating.  Each time, a particle or photon is
diffracted, the amplitude of the grating grows.

In practice, it is difficult or impossible to carry out these experiments with fermions or thermal
atoms.  When we observed superradiance of a condensate, we couldn't observe similar behaviour above
the BEC transition temperature since the threshold laser intensity for superradiant gain is several
orders of magnitude higher (see Ref.\ \cite{inou99super} for details). Furthermore, the
superradiance may be suppressed by heating or other decoherence processes. The shorter coherence
time for non-BEC samples should be even more crucial for the four-wave mixing experiment where the
matter wave grating is probed by very slow atoms which have a long transit time of about 1 ms
through the sample. Another concern are incoherent processes which accompany the stimulated
processes discussed so far. Since the incoherent processes scale linearly with the number of atoms,
whereas the stimulated process is proportional to $N^2$, there is in principle always a regime
where the stimulated process dominates \footnote{In the case of a fermionic sample, the density of
particles can only be increased by increasing the momentum spread of the sample. This can increase
the incoherent ``background'' for four-wave mixing.  For the ensemble in
Fig.~\ref{fig:Momentum_transfer}c an increase in $N$ at constant volume requires the transverse
velocity spread to increase as $N^{1/2}$. Therefore, for large $N$, the incoherent elastic
scattering rate increases as $N^{3/2}$, still more slowly than the stimulated scattering.}.

{\it Discussion}. Coming back to the initial question:  Is matter wave amplification possible for
fermions?  The answer is yes, if the system is prepared in a cooperative state and the
amplification is faster than the coherence time.  However, this amplification does not pile up
atoms in a single quantum state, but rather in states which are in the same (or approximately the
same) momentum state along $\hat z$, but differ in other quantum  numbers. Therefore, this
amplification can be regarded as amplification of a density modulation or as amplification of
spatial bunching of atoms.  Alternatively, one can regard the density modulation as a collective
excitation of the system which involves bosonic quasi-particles (e.g.\ phonons). Superradiance and
four-wave mixing (both with bosons and fermions) can then be ascribed to bosonic stimulation by
those quasi-particles.

The phase-coherent matter wave amplification for fermions would start with a short Bragg pulse
which puts some of the atoms into a recoil state which is then amplified.  This superposition of
two momentum states creates a matter wave grating.  This can be regarded as the interference
pattern of each atom with itself with all the individual interference patterns being exactly in
phase. Matter wave amplification occurs when a single laser beam is diffracted off this grating
increasing the amplitude of each atom to be in the recoiling state. Therefore, the matter wave
amplification scheme of Refs.\ \cite{inou99mwa,kozu99amp} would work for fermions, provided the
whole process can be done in the short coherence time of the fermionic matter wave grating.

Of course, there is a fundamental difference between bosons and fermions which is reflected in the
symmetry of the total wavefunction. A bosonic system with two macroscopically occupied quantum
states is {\it always} in a fully symmetric and maximally cooperative state.  In other words, if
two independent Bose condensates cross each other, there is always a macroscopic interference
pattern (as observed experimentally \cite{andr97int}), which is reflected in $S(q, \omega)$ being
proportional to $N^2$ (or to $N_+ N_-$, to be more precise). It is this density modulation which
can be amplified by the dynamic diffraction discussed in this paper.  If two beams of fermions
overlap, there is no macroscopic interference, unless the two beams were prepared in a symmetric
way, e.g. by generating one of the beams by a Bragg pulse from the other one.

Our discussion of scattering without change of the internal state
can be generalized.  For example, if atoms scatter into the
condensate through a spinflip process, the density grating has to
be replaced by a polarization or coherence grating.  Such
gratings were experimentally studied for laser-cooled atoms
\cite{kuma98}.

This discussion has focused on bosonically enhanced {\it scattering}.  Similarly, bosonic
enhancement of spontaneous emission depends only on a cooperative initial state and not directly
on quantum statistics. For scattering, the relevant coupling strength are the density
fluctuations.  For spontaneous emission, it is the electric dipole moment.  Both are enhanced by
the presence of a Bose condensate, in the latter case because the excited atom corresponds to a
Dicke vector of spin $s=N/2, m=-(N/2)+(1/2)$ which couples more strongly to the vacuum fluctuations
of the electromagnetic field than an individual atom.  Alternatively, the enhanced spontaneous
emission can be regarded as the constructive interference of an ``emitted'' ground state atom with
the macroscopic ground state matter wave.  This picture is analogous to the semi-classical
interpretation of stimulated emission of light.  Ref.\ \cite{sarg74} shows that bosonic
stimulation of photons is due to the constructive interference of the emission of a classical
oscillating dipole with the incident field in the forward direction.

\section{Discussion}

This paper has summarized our recent experiments on Bose-Einstein condensation with the unifying
theme of enhancement and suppression. Suppression of scattering or dissipation can arise for two
different reasons. The phonon and vortex nature of the collective excitations together with energy
and momentum conservation allow dissipation only above a critical velocity. In addition, one has
to consider the dynamics of the excitation process. For microscopic particles, this is reflected
in the matrix element $S(q)$ which characterizes how easily can the condensate absorb momentum in
a scattering process. For macroscopic motion, it is reflected in a critical velocity for vortex
nucleation. Scattering processes are also enhanced by the population of the final states (bosonic
stimulation). Optical stimulation by a laser beam was used in Bragg scattering, and matter wave
stimulation led to superradiance and matter wave amplification.

We have also discussed some subtleties which go beyond the simple
picture using rate equations and occupation numbers. A condensate
in its ground state is in a coherent superposition state of the
zero-momentum state with correlated pairs with momenta $\pm {\bf
q}$ (the quantum depletion). We have shown that the population in
the quantum depletion can cause bosonic stimulation of
spontaneous emission. However, for a scattering situation, there
are two bosonically enhanced pathways which destructively
interfere (causing $S(q) < 1$). Therefore, the concept of bosonic
stimulation can be applied to the many-body state of a
condensate, but with caution. Finally, we have shown that some
form of matter wave amplification is possible in fermionic
samples. This required a careful discussion of bosonic
stimulation by particles vs. quasi-particles and the role of
symmetry vs. quantum degeneracy.

In closing we want to point out that the rich and complex physics displayed here is based
essentially on two four-wave mixing Hamiltonians.  One describes $s$-wave interaction of the
condensate which is responsible for all the many-body effects discussed here including
superfluidity.  The other one couples the atoms to the light (or impurity atoms) and led to
superradiance and amplification of light and atoms.

We are grateful to Alain Aspect, Jean Dalibard,  William D.\ Phillips, Gora Shlyapnikov, and
Philippe Bouyer for organizing a stimulating summer school, to Ananth Chikkatur for contributions
to the sections on collisions, and to Axel G\"{o}rlitz and Aaron Leanhardt for valuable comments on
the manuscript.  This work was supported by NSF, ONR, ARO, NASA, and the David and Lucile Packard
Foundation.

\end{document}